\ifdefined\tmiformat
	\documentclass[journal,twoside,web]{ieeecolor}
\else
	\documentclass[11pt]{article}
\fi

\pdfoutput=1

\ifdefined\tmiformat
	\usepackage{tmi}
	\usepackage{cite}
	\newcommand{\citep}{\cite}
\else
	\usepackage{natbib}
	\usepackage{authblk}
	\usepackage{fullpage}
	\usepackage[labelfont=bf]{caption}
	\usepackage{times}
	\usepackage[export]{adjustbox}
\fi
\usepackage{amsmath,amssymb,amsfonts}
\usepackage{graphicx}
\usepackage{anyfontsize}
\usepackage{balance}

\usepackage{overpic}
\usepackage{mathtools}
\usepackage{algorithm}
\usepackage{soul}
\usepackage{relsize}
\usepackage{algpseudocode}
\usepackage[customcolors]{hf-tikz}
\usepackage[hyperfootnotes=false]{hyperref}
\usepackage{fancyhdr}

\newlength{\hcolw}
\newcommand{\pamine}{}
\newcommand{\pbmine}{}
\newcommand{\auxminebis}{}
\newcommand{\capsepmine}{}
\newcommand{\figstaastmine}{}
\newcommand{\figendastmine}{}
\newcommand{\figstamine}{}
\newcommand{\figendmine}{}
\newcommand{\tabstaastmine}{}
\newcommand{\tabendastmine}{}

\ifdefined\tmiformat
	
	\renewcommand{\auxminebis}{\fontsize{8.5}{11}}
	\renewcommand{\pbmine}{-4}
	
	\renewcommand{\capsepmine}{\vspace{0mm}}
	\renewcommand{\figstamine}{\begin{figure}[!t]}
	\renewcommand{\figendmine}{\end{figure}}
	\renewcommand{\figstaastmine}{\begin{figure*}[!t]}
	\renewcommand{\figendastmine}{\end{figure*}}
	\renewcommand{\tabstaastmine}{\begin{table*}[!t]}
	\renewcommand{\tabendastmine}{\end{table*}}
\else
	
	\renewcommand{\auxminebis}{}
	\renewcommand{\pbmine}{-6}
	
	\renewcommand{\capsepmine}{\vspace{2mm}}
	\renewcommand{\figstamine}{\begin{figure}[!htb]}
	\renewcommand{\figendmine}{\end{figure}}
	\renewcommand{\figstaastmine}{\begin{figure}[!htb]}
	\renewcommand{\figendastmine}{\end{figure}}
	\renewcommand{\tabstaastmine}{\begin{table}[!htb]}
	\renewcommand{\tabendastmine}{\end{table}}
\fi

\usetikzlibrary{shapes,arrows}
\tikzstyle{block} = [draw, fill=cyan!20, rectangle, rounded corners=0.5cm,
    minimum height=3em, minimum width=5em]
\tikzstyle{input} = [coordinate]

\pdfpageattr {/Group << /S /Transparency /I true /CS /DeviceRGB>>}

\newcommand\argmin{\ensuremath{\operatornamewithlimits{argmin}}}
\newcommand\median{\ensuremath{\operatornamewithlimits{median}}}

\newcommand\id{\ensuremath{\operatornamewithlimits{id}}}

\definecolor{naranja}{rgb}{1.0,0.5,0}
\definecolor{violeta}{rgb}{0.5,0,1}
\definecolor{verde}{rgb}{0.75,1,0.75}
\definecolor{rojo}{rgb}{1,0.75,0.75}
\definecolor{azul}{rgb}{0.75,0.75,1}
\definecolor{naranjal}{rgb}{1.0,0.85,0.7}

\tikzstyle{nosep}=[inner sep=0pt, outer sep=0pt]

\DeclarePairedDelimiter\ceil{\lceil}{\rceil}

\ifdefined\tmiformat

	\def\BibTeX{{\rm B\kern-.05em{\sc i\kern-.025em b}\kern-.08em
    	T\kern-.1667em\lower.7ex\hbox{E}\kern-.125emX}}
\begin{document}
		\title{Fetal MRI by robust deep generative prior reconstruction and diffeomorphic registration: application to gestational age prediction}
		\author{Lucilio~Cordero-Grande, Juan~Enrique~Ortu{\~n}o-Fisac, Alena~Uus, Maria~Deprez, Andr{\'e}s~Santos, \IEEEmembership{Senior Member, IEEE}, Joseph~V.~Hajnal, Mar{\'i}a~Jes{\'u}s~Ledesma-Carbayo, \IEEEmembership{Senior Member, IEEE}
		\thanks{Date of submission: 2021/06/31. This work is funded by the Ministry of Science and Innovation, Spain, under the Beatriz Galindo Programme [BGP18/00178]. This work has been supported by the Madrid Government (Comunidad de Madrid-Spain) under the Multiannual Agreement with Universidad Polit\'ecnica de Madrid in the line Support for R\&D projects for Beatriz Galindo researchers, in the context of the V PRICIT (Regional Programme of Research and Technological Innovation).}
		\thanks{L. C. G., J. E. O. F., A. S., and M. J. L. C. are with the Biomedical Image Technologies, ETSI Telecomunicaci\'on, Universidad Polit\'ecnica de Madrid, Madrid, Spain. (e-mail: lucilio.cordero@upm.es)}
		\thanks{L. C. G., J. E. O. F., A. S., and M. J. L. C. are with the Biomedical Research Networking Center in Bioengineering, Biomaterials and Nanomedicine (CIBER-BBN), Madrid, Spain.}
		\thanks{L. C. G., A. U., M. D. and J. V. H are with the Centre for the Developing Brain and Biomedical Engineering Department, School of Biomedical Engineering and Imaging Sciences, King's College London, King's Health Partners, St Thomas' Hospital, London, SE1 7EH, UK}}	
	
\else

	\hyphenpenalty=500
	\providecommand{\keywords}[1]{\small{\textbf{\textit{Index terms---}} #1}}
	\hypersetup{
    	pdftitle={Fetal MRI by robust deep generative prior reconstruction and diffeomorphic registration: application to gestational age prediction},    
	    pdfauthor={Lucilio Cordero-Grande},     
    	pdfkeywords={A, B, C, D, E}, 
	}	

	\title{Fetal MRI by robust deep generative prior reconstruction and diffeomorphic registration: application to gestational age prediction}
	\date{}
	\author[1,2,3]{Lucilio~Cordero-Grande}
	\author[1,2]{Juan~Enrique~Ortu{\~n}o-Fisac}
	\author[3]{Alena~Uus}
	\author[3]{Maria~Deprez}
	\author[1,2]{Andr{\'e}s~Santos}
	\author[3]{Joseph~V.~Hajnal}
	\author[1,2]{Mar{\'i}a~Jes{\'u}s~Ledesma-Carbayo}
	\affil[1]{Biomedical Image Technologies, ETSI Telecomunicaci\'on, Universidad Polit\'ecnica de Madrid, Madrid, Spain}
	\affil[2]{Biomedical Research Networking Center in Bioengineering, Biomaterials and Nanomedicine (CIBER-BBN), Madrid, Spain}
	\affil[3]{Centre for the Developing Brain and Biomedical Engineering Department\\School of Biomedical Engineering and Imaging Sciences\\King's College London, King's Health Partners, St Thomas' Hospital, London, SE1 7EH, UK\vspace{2mm}}
	\affil[ ]{\small{lucilio.cordero@upm.es, juanen@die.upm.es, alena.uus@kcl.ac.uk, maria.deprez@kcl.ac.uk, andres@die.upm.es, jo.hajnal@kcl.ac.uk, mledesma@die.upm.es}}
	\begin{document}
\fi

\maketitle

\begin{abstract}

Magnetic resonance imaging of whole fetal body and placenta is limited by different sources of motion affecting the womb. Usual scanning techniques employ single-shot multi-slice sequences where anatomical information in different slices may be subject to different deformations, contrast variations or artifacts. Volumetric reconstruction formulations have been proposed to correct for these factors, but they must accommodate a non-homogeneous and non-isotropic sampling, so regularization becomes necessary. Thus, in this paper we propose a deep generative prior for robust volumetric reconstructions integrated with a diffeomorphic volume to slice registration method. Experiments are performed to validate our contributions and compare with a state of the art method in a cohort of $72$ fetal datasets in the range of $20$-$36$ weeks gestational age. Results suggest improved image resolution and more accurate prediction of gestational age at scan when comparing to a state of the art reconstruction method. In addition, gestational age prediction results from our volumetric reconstructions compare favourably with existing brain-based approaches, with boosted accuracy when integrating information of organs other than the brain. Namely, a mean absolute error of $0.618$ weeks ($R^2=0.958$) is achieved when combining fetal brain and trunk information.

\end{abstract}


\section{Introduction}

\label{sec:INTR}

\ifdefined\tmiformat
	\IEEEPARstart{M}{agnetic}
\else
	Magnetic 
\fi
Resonance Imaging (MRI) is indicated when both Central Nervous System (CNS) and non CNS fetal anomalies are suspected on ultrasound~\citep{ACR-SPR20,Herrera20}. When compared to ultrasound, MRI is a unique instrumental technique for studying fetal development due to enlarged Field Of View (FOV), superior soft tissue contrast, and lack of shadowing. Basic examination protocols involve collecting $T_2$-weighted slices along the main axes of the fetal organs. In the case of brain imaging, aspirations for more quantitative imaging motivated the development of Slice to Volume (SV) reconstruction techniques aiming to obtain a volumetric representation of the fetal brain from the set of collected slices~\citep{Gholipour10,Kuklisova-Murgasova12}. If SV reconstruction is available, acquisitions may be redesigned for collecting the most diverse and efficient set of orientations considering both scanning limitations and reconstruction properties. Free reformatting of the imaging plane may be even more important for non-brain applications, where, due to motion, limited resolution, Signal to Noise Ratio (SNR), and scanning time, obtaining the principal axes of various target organs while planning the scans may be difficult or infeasible. However, reconstruction of the whole fetus, uterus and placenta has encountered stronger challenges when compared with brain only reconstructions due to increased complexity of different non-rigid sources of motion.

Despite non-rigid motion transformation models for SV registration problems have been proposed some time ago~\citep{Ferrante17}, the predominance of fetal brain applications has delayed their incorporation to whole body fetal image reconstruction algorithms. After some preliminary works applying rigid correction models outside the brain,~\cite{Alansary17} proposed a method using patch-based rigid registration to approximate non-rigid deformations. Improved performance was shown in~\cite{Uus20a} when using deformable registration based on free-form deformations on a hierarchical B-Spline grid, reconstruction based on weighted Gaussian interpolation, bias correction, and global and local outlier rejection based on normalized cross correlation and structural similarity indexes respectively. As reported in~\cite{Ferrante17}, other explored alternatives for deformable motion models in SV registration include usage of thin plate splines and finite element meshes.

On the other hand, application of Deep Learning (DL) methodologies in the orbit of fetal SV reconstruction includes methods for rigid alignment of slices to brain templates~\citep{Hou18,Salehi19}, rigid motion tracking~\citep{Singh20}, automatic localization of the fetal brain~\citep{Ebner20}, or image quality assessment~\citep{Largent21}. Distinctly, in this work we focus on the integration of DL architectures within the reconstruction formulation. Most efforts on DL for inverse problems have focused on learning a mapping between an approximate inverse of a fully characterized measurement operator and a ground truth reconstruction, which is used at test time to mitigate spurious residuals in the reconstruction, typically by unrolled schemes~\citep{Ongie20}. However, generation of ground truth reconstructions is problematic in our application as it would involve the acquisition of oversampled datasets in a sensible population. In addition, as the measurements are affected by motion, learning may be biased by the reconstruction method employed to build the training data.

Difficulties with ground truth are also an issue for validation, differences between reconstruction methods are often subtle or difficult to summarize, and improved reconstructions may not necessarily impact a particular clinical application. For these reasons, we have combined generic measures of data quality with a task-based validation strategy based on assessing the Gestational Age (GA) prediction performance. This is identified as a clinically relevant task because GA knowledge is critical for fetal development characterization from imaging and errors or uncertainties in GA predictions could be indicative of developmental abnormalities~\citep{Shi20}.

In this work we propose the Robust Generative Diffeomorphic Slice to Volume Reconstruction (RGDSVR) method. Our contributions include an efficient version of the Large Deformation Diffeomorphic Metric Mapping (LDDMM)~\citep{Faisal05} framework adapted to the computational requirements of joint deformable registration and reconstruction problems. In addition, we propose a robust explicit inverse formulation of the reconstruction that makes use of an untrained generative model, the so-called Deep Decoder (DD)~\citep{Heckel19}, for regularization. Finally, we show that free reformatting of whole body fetal SV reconstructions can be leveraged for accurate estimates of GA at scan. The source code and exemplary data required to reproduce the main results of the paper is made available at \url{https://github.com/lcorgra/RGDSVR/releases/tag/1.0.0}.

\section{Methods}

\label{sec:METH}

Common structural fetal MRI protocols are based on the acquisition of a series of stacks of single-shot slices along different orientations (see Fig.~\ref{fig:FIG00}). Thus, we start from a data array $\mathbf{y}=(y_{m^l_1m^l_2m^l_3l})$ encompassing several stacks $1\leq l\leq N_l$, where $m^l_1m^l_2m^l_3$ indexes a voxel in stack $l$, respectively along the readout, phase encode and slice directions, and $N_l$ denotes the number of stacks. Then, we formulate the reconstruction problem as the recovery of a volume $\mathbf{x}$ from $\mathbf{y}$ via
\begin{equation}
\label{eq:FORM}
(\hat{\mathbf{x}},\hat{\boldsymbol{\phi}},\hat{\boldsymbol{\theta}})=\argmin_{\mathbf{x},\boldsymbol{\phi},\boldsymbol{\theta}}f(\mathbf{A}\boldsymbol{\Phi}^{\boldsymbol{\phi}}\mathbf{x}-\mathbf{y})+g(\mathbf{x}-\boldsymbol{\Theta}^{\boldsymbol{\theta}}\mathbf{z})+h(\boldsymbol{\phi}),
\end{equation}
with $f$ a robust loss function, $g$ a reconstruction regularizer, and $h$ a motion regularizer. In this formulation, we model the image formation by a measurement operator $\mathbf{A}$, the fetal motion by a diffeomorphic image warping operator $\boldsymbol{\Phi}^{\boldsymbol{\phi}}$ with deformation described by the parameter vector $\boldsymbol{\phi}$, and use a generative DD network operator $\boldsymbol{\Theta}^{\boldsymbol{\theta}}$ with learnable parameters $\boldsymbol{\theta}$ and fixed random input $\mathbf{z}$ for regularization.

\figstaastmine
\centerline{\begin{overpic}[width=\textwidth]{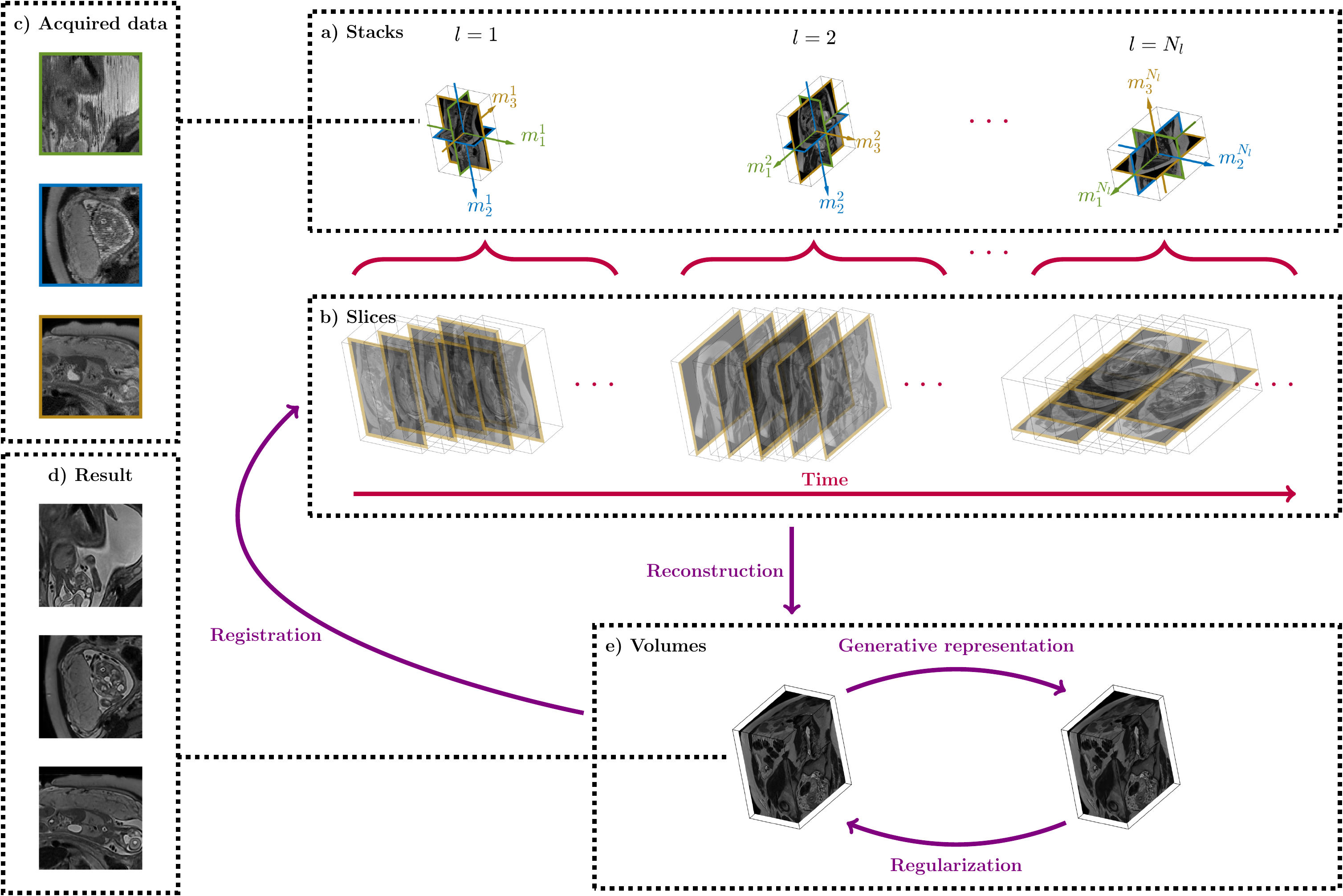}\end{overpic}}
\caption{Overview of SV reconstruction and registration for whole body fetal imaging. \textbf{a)} Stacks of slices are acquired along different orientations, here with axial, sagittal and coronal examples. \textbf{b)} Slices collection follows different interleaving configurations in time.  \textbf{c)} Acquired data may include corrupted slices and appear discontinuous in the slice direction due to motion. \textbf{d)} Volumetric reconstructions achieving a dense representation of the fetal anatomy can be obtained by accounting for these measurement factors. \textbf{e)} For this sake, the algorithm alternates between robust motion compensated SV reconstruction, multi-scale registration of reconstructed volume to acquired slices for motion refinement, projection of the reconstruction solution to a natural image generative representation, and regularization of the reconstruction using this projection.}
\label{fig:FIG00}
\figendastmine

In~\S~\ref{sec:LOSS} we describe the robust loss function $f$, in~\S~\ref{sec:MEAS} the measurement operator $\mathbf{A}$, in~\S~\ref{sec:MUME} the temporal structure of the acquisition, in~\S~\ref{sec:MOTI} the registration methodology to obtain $\boldsymbol{\Phi}^{\boldsymbol{\phi}}$, in~\S~\ref{sec:NETW} the deep generative prior $\boldsymbol{\Theta}^{\boldsymbol{\theta}}\mathbf{z}$ used as a regularizer, in~\S~\ref{sec:MATE} the fetal MRI cohort used to test our reconstructions, in~\S~\ref{sec:IMDE} the adopted reconstruction implementation, and in~\S~\ref{sec:GAES} the GA estimation procedure.

\subsection{Robust reconstruction}

\label{sec:LOSS}

The residuals of the reconstruction are denoted by
\begin{equation}
\label{eq:RESI}
\mathbf{r}=\mathbf{A}\boldsymbol{\Phi}\mathbf{x}-\mathbf{y}=(r_{m_1^lm_2^lm_3^ll})
\end{equation}
The corresponding squared residuals map is denoted by $\mathbf{r}^2=(r_{m_1^lm_2^lm_3^ll}^2)$. We use a smooth version of the Welsch metric due to its strong robustness to outliers~\citep{Holland77}:
\begin{equation}
\label{eq:LORE}
f(\mathbf{r})=\frac{\kappa^2\sigma^2}{2}\left\|1-\exp\left(-\frac{\boldsymbol{\mathcal{G}}\mathbf{r}^2}{\kappa^2\sigma^2}\right)\right\|_1,
\end{equation}
with $\boldsymbol{\mathcal{G}}$ an in-plane Gaussian smoothing operator to weight the contribution of the observation in a voxel according to the residuals in its neighborhood, $\sigma$ a robust estimation of scale, $\sigma^2=2.1981\median(\boldsymbol{\mathcal{G}}\mathbf{r}^2)$~\citep{Holland77}, and $\kappa$ a tuning constant.

We can solve for $\mathbf{x}$ using Iteratively ReWeighted Least Squares (IRWLS)~\citep{Holland77,Maronna19}. Defining the encoding operator as $\mathbf{E}=\mathbf{A}\boldsymbol{\Phi}$, the update at iteration $i+1$, which can be computed via conjugate gradient, is
\begin{equation}
\hat{\mathbf{x}}_{(i+1)}=(\mathbf{E}^T\mathbf{W}_{(i)}\mathbf{E})^{-1}\mathbf{E}^T\mathbf{W}_{(i)}\mathbf{y},
\label{eq:IRLS}
\end{equation}
with $\mathbf{W}_{(i)}$ a diagonal matrix of weights. The diagonal entries for the weights, $\mathbf{w}_{(i)}$, are obtained from the squared residuals and the updated scale at iteration $i$, $\mathbf{r}^2_{(i)}$ and $\sigma^2_{(i)}$ as~\citep{Holland77,Maronna19}:
\begin{equation}
\mathbf{w}_{(i)}=\boldsymbol{\mathcal{G}}\exp\left(-\frac{\boldsymbol{\mathcal{G}}\mathbf{r}^2_{(i)}}{\kappa^2\sigma^2_{(i)}}\right).
\label{eq:WEEQ}
\end{equation}
To improve the IRWLS convergence we use homotopy continuation~\citep{Sidney94}. The computed weights are modified for reconstruction according to:
\begin{equation}
\label{eq:COSH}
\begin{split}
&\mathbf{W}_{(i)}=\text{diag}(\mathbf{w}_{(i)})^{\tau_{(i)}},\\
&\tau_{(0)}=0,\quad\tau_{(i+1)}=\min((\tau_{(i)}+1)/N_i,1),
\end{split}
\end{equation}
so an Ordinary Least Squares (OLS) problem is solved in the first iteration ($\tau=0$, $\mathbf{W}=\mathbf{I}$, the identity matrix) and the target formulation is gradually reached after $N_i$ iterations ($\tau=1$). In Fig.~\ref{fig:FIG01} we show an example including a sampled slice, residuals at first and last iterations of the reconstruction and corresponding weights. 

\figstaastmine
\setlength{\hcolw}{0.19\textwidth}
\renewcommand{\pamine}{29}
\centerline{
\begin{overpic}[width=\hcolw,trim={0 29mm 0 13mm},clip]{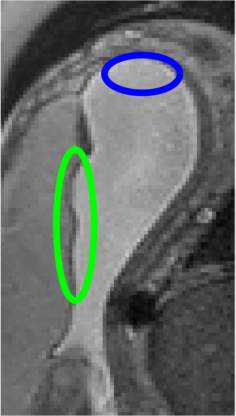}\put(\pamine,\pbmine)
{\makebox(-50,0){\auxminebis\textcolor{black}{\textbf{a)}}}}\end{overpic}\hspace{1mm}
\begin{overpic}[width=\hcolw,trim={0 29mm 0 13mm},clip]{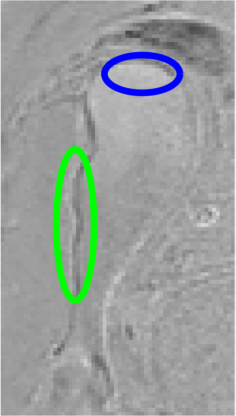}\put(\pamine,\pbmine)
{\makebox(-50,0){\auxminebis\textcolor{black}{\textbf{b)}}}}\end{overpic}\hspace{1mm}
\begin{overpic}[width=\hcolw,trim={0 29mm 0 13mm},clip]{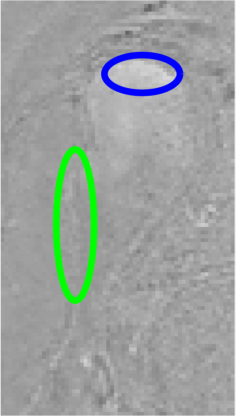}\put(\pamine,\pbmine)
{\makebox(-50,0){\auxminebis\textcolor{black}{\textbf{c)}}}}\end{overpic}\hspace{1mm}
\begin{overpic}[width=\hcolw,trim={0 29mm 0 13mm},clip]{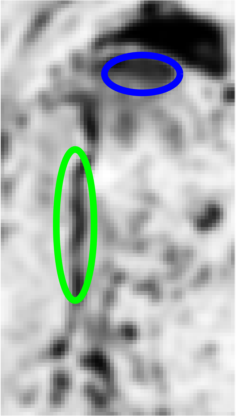}\put(\pamine,\pbmine)
{\makebox(-50,0){\auxminebis\textcolor{black}{\textbf{d)}}}}\end{overpic}\hspace{1mm}
\begin{overpic}[width=\hcolw,trim={0 29mm 0 13mm},clip]{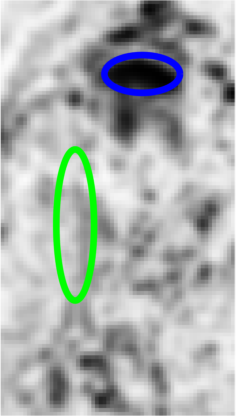}\put(\pamine,\pbmine)
{\makebox(-50,0){\auxminebis\textcolor{black}{\textbf{e)}}}}\end{overpic}
}
\capsepmine
\caption{\textbf{a)} Measured data $\mathbf{y}$, residuals $\mathbf{r}$ at \textbf{b)} first and \textbf{c)} last iterations of the reconstruction, and weights $\mathbf{w}$ at \textbf{d)} first and \textbf{e)} last iterations of the reconstruction. The green ellipse encloses an area around the the placenta and amniotic sac boundary with high residuals when starting the reconstruction --see \textbf{b)}--. These become smaller at last iteration --see \textbf{c)}-- due to corrected motion, which increases the reliability of the data in that area, as shown by the corresponding weights. The blue ellipse encloses an area where flow artifacts cause a relative enhancement of the amniotic fluid so, driven by anomalously high residuals, weights are kept low in the last iteration --see \textbf{e)}--.}
\label{fig:FIG01}
\figendastmine

\subsection{Measurement operator}

\label{sec:MEAS}

The measurement operator $\mathbf{A}$, is defined for each stack $l$ as $\mathbf{A}_l=\boldsymbol{\mathcal{D}}_l\mathbf{T}_l$, where $\mathbf{T}_l$ is a rigid transformation accounting for the orientation of the stack $l$ and $\boldsymbol{\mathcal{D}}_l$ is the slice sampling operator. We model blurring and discretization in $\boldsymbol{\mathcal{D}}_l$ using a Gaussian slice profile and according to the slice thickness and slice separation of stack $l$~\citep{Noll97}. Modeling and estimation of image inhomogeneities had a small impact in the tested data, so it is left out of this manuscript. However, inhomogeneities may become significant at higher field strengths.

\subsection{Temporal structure of the scan}

\label{sec:MUME}

Multi-slice scans are typically collected in a non-sequential manner where consecutively acquired slices are located distant to each other. An example of the particular slice order used in a given stack of the acquisition is shown in Fig.~\ref{fig:FIG02}a. We define an interleave as a series of slices acquired within a single FOV sweep, so in this example we have $20$ interleaves. As illustrated in Fig.~\ref{fig:FIG02}b, we can also group our slices into so-called packages, constructed by considering the gap between the first slices in two consecutive sweeps, with $4$ packages per stack in our sequence.

For motion estimation we define an operator $\mathbf{P}_s^{(j)}$, $1\leq s\leq N_s^j$, to extract the slices associated to the motion state $s$. Motion states are constructed to correspond to a given stack, package or interleave depending on the multi-scale motion estimation level $j\in\{\texttt{STACK,PACKAGE,INTERLEAVE}\}$ at which we are operating, with $N_s^j$ the number of states at level $j$. We start by estimating a deformation per stack. Then, we propagate these estimates as starting transformations for estimations at the package level. Finally, we propagate the latter to estimate a deformation per-interleave, which serves to accommodate different slice deformations because subsequent slices are acquired with a substantial gap.

\figstamine
\renewcommand{\pamine}{50}\vspace{2mm}
\centerline{
\begin{overpic}[width=0.49\textwidth]{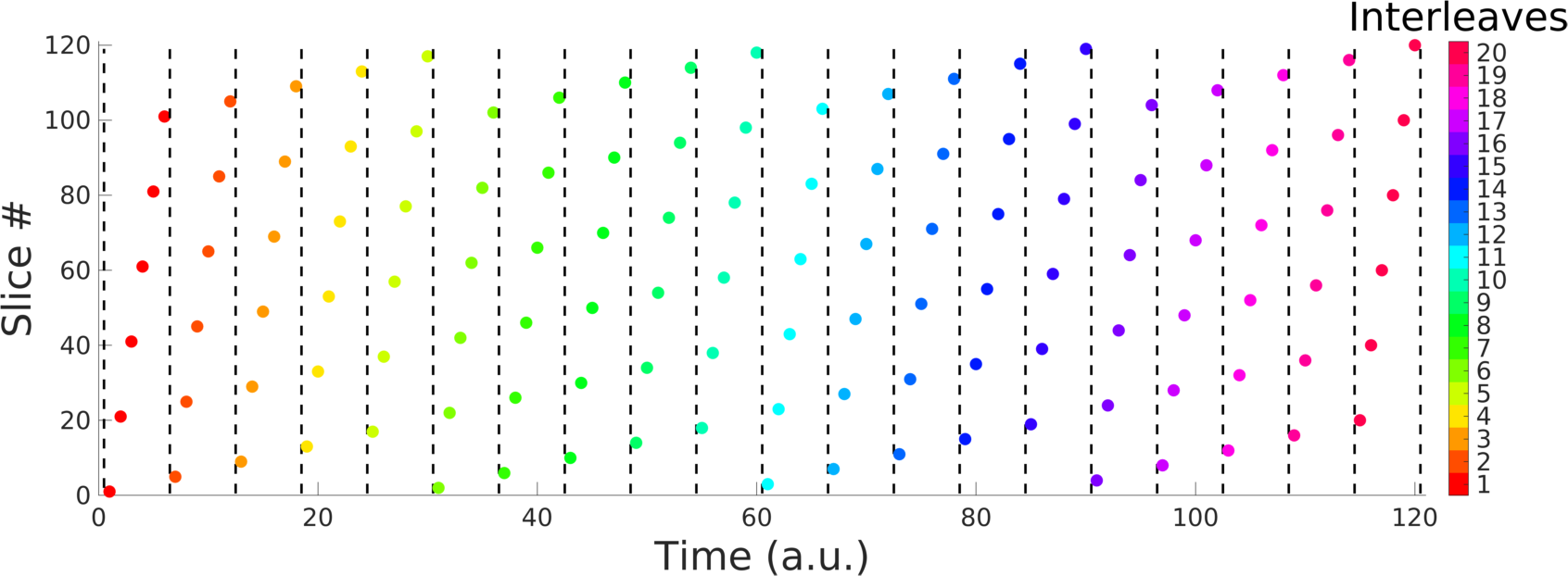}\put(\pamine,\pbmine)
{\makebox(-75,10){\auxminebis\textbf{a)}}}\end{overpic}
\ifdefined\tmiformat
	}\vspace{2mm}\centerline{
\fi
\begin{overpic}[width=0.49\textwidth]{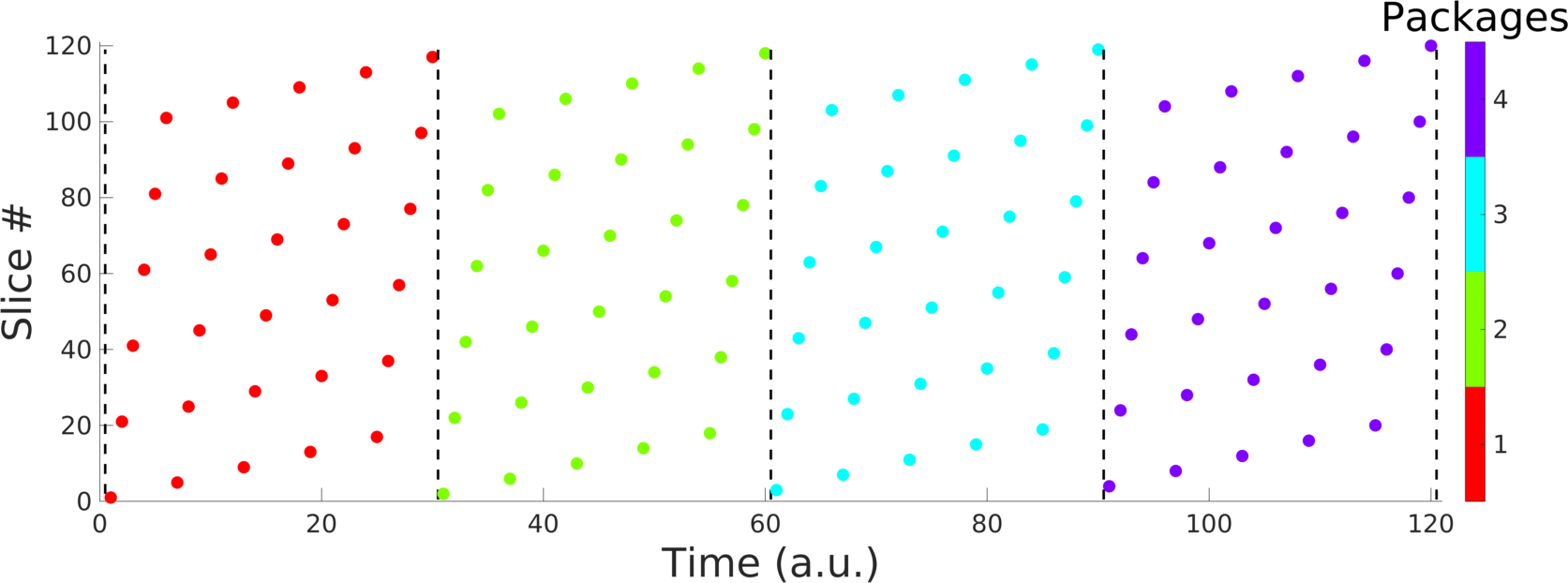}\put(\pamine,\pbmine)
{\makebox(-75,10){\auxminebis\textbf{b)}}}\end{overpic}}
\capsepmine
\caption{Slice acquisition order for an exemplary stack highlighting \textbf{a)} interleave and \textbf{b)} package structure.}
\label{fig:FIG02}
\figendmine

\subsection{Diffeomorphic registration}

\label{sec:MOTI}

In the LDDMM framework~\citep{Faisal05} a transformation between two images $\varphi_1$ is given as the end point of the flow of a vector field $v_t(\varphi_t)=\varphi_t'$, where $\varphi_0=\id$ with $\id$ the identity function. The diffeomorphic registration between the source and target volumes $I_0$ and $I_1$ is posed as the variational minimization of:
\begin{equation}
\label{eq:DIRE}
E(v)=\frac{1}{\sigma_{\text{I}}^2}\|I_0(\varphi_1^{-1})-I_1\|_{L_2}^2+\int_0^1\|v_t\|_V^2dt,
\end{equation} 
where $\sigma_{\text{I}}$ represents the image noise variance and norms are taken in the spaces of square integrable functions $L_2$ and allowed fields $V$. Then, by enforcing a certain smoothness on $V$, $\varphi_t$ is guaranteed to lie in the space of diffeomorphisms. In practice, smoothness is promoted by an operator $\mathcal{L}=(-\beta\Delta+\id)^\gamma$ with $\Delta$ the Laplacian and $\beta$ and $\gamma$ controlling the level and properties of smoothness respectively.

Ensuring smooth and invertible mappings by the LDDMM framework may confront numerical difficulties~\cite{Mang17}. For efficiency, we adopt single-step temporal integration and track the invertibility based on the sufficient Lipschitz condition~\cite{Chen08}:
\begin{equation}
\label{eq:LIJA}
q(p)=\max_{p'}\frac{\|(\varphi(p)-\varphi(p'))-(p-p')\|_2^2}{\|p-p'\|_2^2}<a,
\end{equation}
with $p$, $p'$ a pair of coordinates and Lipschitz constant $a<1$. A risk of local non-invertibility (with maximum for $p'$ taken in a discretized neighborhood of $p$) is used in each step to modulate the registration gradient descent step size by $\max(a-\max_p(q(p)),0)$.

Hence, in our joint SV reconstruction and diffeomorphic registration formulation, we face a series of registration subproblems from source reconstructions to target measurements:
\begin{equation}
\label{eq:DDRR}
\hat{\boldsymbol{\phi}}_s^{(j)}=\argmin_{\boldsymbol{\phi}_s^{(j)}}\frac{1}{\sigma_{\text{I}}^2}\|\mathbf{P}_s^{(j)}(\mathbf{A}\boldsymbol{\Phi}^{\boldsymbol{\phi}_s^{(j)}}\mathbf{x}-\mathbf{y})\|_2^2+(\boldsymbol{\mathcal{L}}\boldsymbol{\phi}_s^{(j)})^T\boldsymbol{\phi}_s^{(j)}.
\end{equation}
$\boldsymbol{\Phi}^{\boldsymbol{\phi}_s^{(j)}}$ is the image warping operator, implemented by linear interpolation according to the diffeomorphism induced by $\boldsymbol{\phi}_s^{(j)}$, which parametrizes a band-limited field~\cite{Zhang19}. $\boldsymbol{\mathcal{L}}$ represents the discrete version of the smoothness operator $\mathcal{L}$.~\eqref{eq:DDRR} corresponds to the optimization with respect to the motion parameters of~\eqref{eq:FORM} with $\tau=0$ in~\eqref{eq:COSH} and regularization
\begin{equation}
\label{eq:MORE}
h(\boldsymbol{\phi}^{(j)})=\sigma_{\text{I}}^2\sum_{s}(\boldsymbol{\mathcal{L}}\boldsymbol{\phi}_s^{(j)})^T\boldsymbol{\phi}_s^{(j)}.
\end{equation}
Although the robust cost function presented in~\S\ref{sec:LOSS} could also be used for motion estimation, we have observed that OLS is effective in escaping local optima of the motion parameters. The Hilbert gradient in the space of diffeomorphisms $V$~\citep{Faisal05} is obtained by:
\begin{equation}
\label{eq:GRAM}
(\nabla_{\boldsymbol{\phi}_s^{(j)}}E)_V=\boldsymbol{\mathcal{L}}^{-1}\left(\frac{2}{\sigma_{\text{I}}^2}(\nabla\boldsymbol{\Phi}^{\boldsymbol{\phi}_s^{(j)}}\mathbf{x})\cdot(\mathbf{A}^T\mathbf{P}_s^{(j)}\mathbf{r})\right)+2\boldsymbol{\phi}_s^{(j)}.
\end{equation}

\subsection{DD regularization}

\label{sec:NETW}

The DD~\citep{Heckel19} is a deep architecture designed for the efficient representation of natural images without using training data. The proposed network is based on the concatenation of upsampling, convolution, activation and batch normalization layers where $1\times 1$ convolutions (i.e., linear layers) are shown to be the most cost-effective. Its parameters are optimized by fitting to a particular cost function using fixed random inputs $\mathbf{z}$, and it has shown competitive results when compared to supervised methods in applications such as image compression, denoising, inpainting, and reconstruction~\citep{Zalgabi20}.

The proposed DD architecture is depicted in Fig.~\ref{fig:FIG03}. We have performed a series of modifications to the original architecture. First, swish units are used instead of rectified linear units, according to the results in~\cite{Ramachandran17}. Second, batch normalization is applied before rather than after activation, following the recommendations in~\cite{Ioffe15}. Third, we use sinc rather than bilinear upsampling to better preserve fine detailed structures. Finally, we parametrize our architecture using the number of scales $D$, the scaling ratio $U$, the channel compression ratio per-scale $C$ (defined as the ratio of input and output features of the linear layers), and the number of output channels of the first linear layer at full resolution $K$. Similar to~\cite{Heckel19}, we empirically fix $D=5$ and $U=2$, and $K$ is determined by prescribing a given minimum compression rate $S$ for the network. However, we use $C=3$ instead of $C=1$, as this provides a thinner network at the finest scale, which reduces peak memory consumption, a limiting factor in 3D.

\figstaastmine
\centerline{
\begin{overpic}[width=\textwidth]{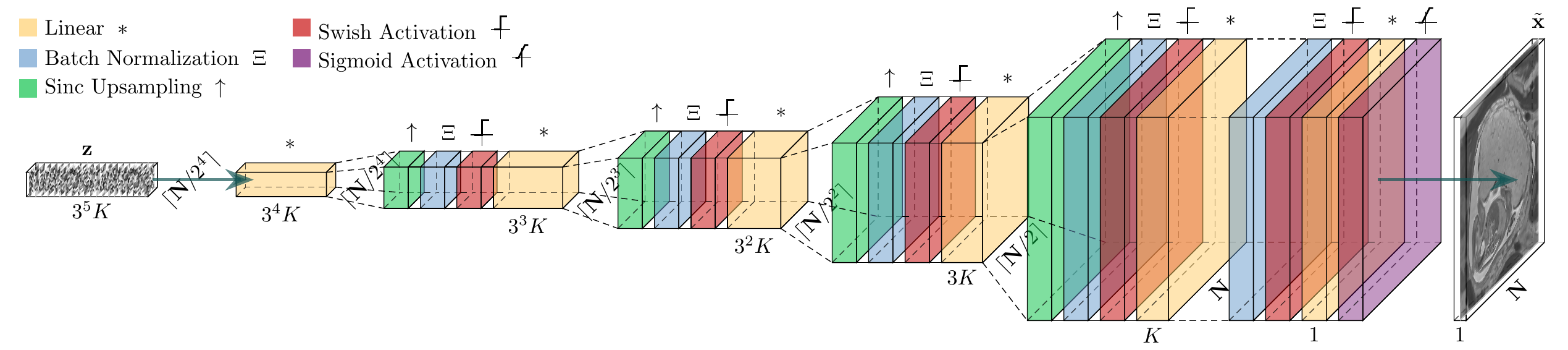}\end{overpic}
}
\caption{DD architecture used for reconstruction regularization. Considering a network with $D$ scales, scaling ratio $U$ and channel compression ratio per-scale $C$, the input is a random array of size $\ceil{\mathbf{N}/U^{D-1}}\times C^DK$, with $\mathbf{N}=(N_1,N_2,N_3)$ the grid dimensions of the output image, $K$ the number of channels at the finest scale, and $\ceil{\cdot}$ the ceiling function. The coarsest scale comprises simply a linear layer. The remaining scales are connected by sinc upsampling layers and formed by blocks of batch normalization, activation and linear layers. At the finest scale, we use two such blocks followed by sigmoid activation. Linear layers at scale $0\leq d<D$ map $C^{d+1}K$ input to $C^dK$ output features but for the additional layer at full resolution that performs the final $K\rightarrow 1$ mapping.}
\label{fig:FIG03}
\figendastmine

The DD output is used as the expected value of the reconstruction for the following generalized Tikhonov regularizer:
\begin{equation}
g(\mathbf{x}-\boldsymbol{\Theta}^{\boldsymbol{\theta}}\mathbf{z})=\lambda\|\mathbf{x}-\boldsymbol{\Theta}^{\boldsymbol{\theta}}\mathbf{z}\|_2^2.
\label{eq:PRIM}
\end{equation}
For a series of fixed parameters of the network at reconstruction iteration $i$, $\hat{\boldsymbol{\theta}}_{(i)}$, we can compute the DD output
\begin{equation}
\tilde{\mathbf{x}}_{(i)}=\boldsymbol{\Theta}^{\hat{\boldsymbol{\theta}}_{(i)}}\mathbf{z}
\label{eq:NEOU}
\end{equation}
mapped to match the dynamic range of the reconstruction at previous iteration, and solve the reconstruction problem
\begin{equation}
\hat{\mathbf{x}}_{(i+1)}=\argmin_{\mathbf{x}}f(\mathbf{A}\boldsymbol{\Phi}\mathbf{x}-\mathbf{y})+\lambda\|\mathbf{x}-\tilde{\mathbf{x}}_{(i)}\|_2^2
\label{eq:RECP}
\end{equation}
by extending the solution in~\eqref{eq:IRLS} with the Tikhonov term. Then, the parameters of the network can be refined according to:
\begin{equation}
\label{eq:FIWU}
\hat{\boldsymbol{\theta}}_{(i+1)}=\argmin_{\boldsymbol{\theta}}\|\boldsymbol{\Theta}^{\boldsymbol{\theta}}\mathbf{z}-\hat{\mathbf{x}}_{(i+1)}\|_2^2.
\end{equation}
This scheme penalizes the deviation of the reconstruction from the space of natural images, a relaxed version of the formulation in~\cite{Heckel19,Zalgabi20}.

\subsection{Materials}

\label{sec:MATE}

We have tested our reconstruction algorithm in a cohort of $72$ fetal cases consented as participants in the iFIND project (ISRCTN16542843)~\citep{iFIND} with GA distribution ranging from $20$+$2$ to $36$+$0$ weeks shown in Fig.~\ref{fig:FIG07}a. The cohort includes controls, fetuses with suspected abnormalities, and cases with incidental findings. In Fig.~\ref{fig:FIG07}b we show the distribution of abnormalities, including controls or cases without any detected anomaly; with renal, urinary or genital abnormalities; with gastrointestinal abnormalities including abdominal wall defects; with chest abnormalities including respiratory, cardiac and thoracic; and with multiple abnormalities including skeletal and CNS anomalies.

\figstamine
\renewcommand{\pamine}{50}\vspace{2mm}
\centerline{
\begin{overpic}[width=0.49\textwidth]{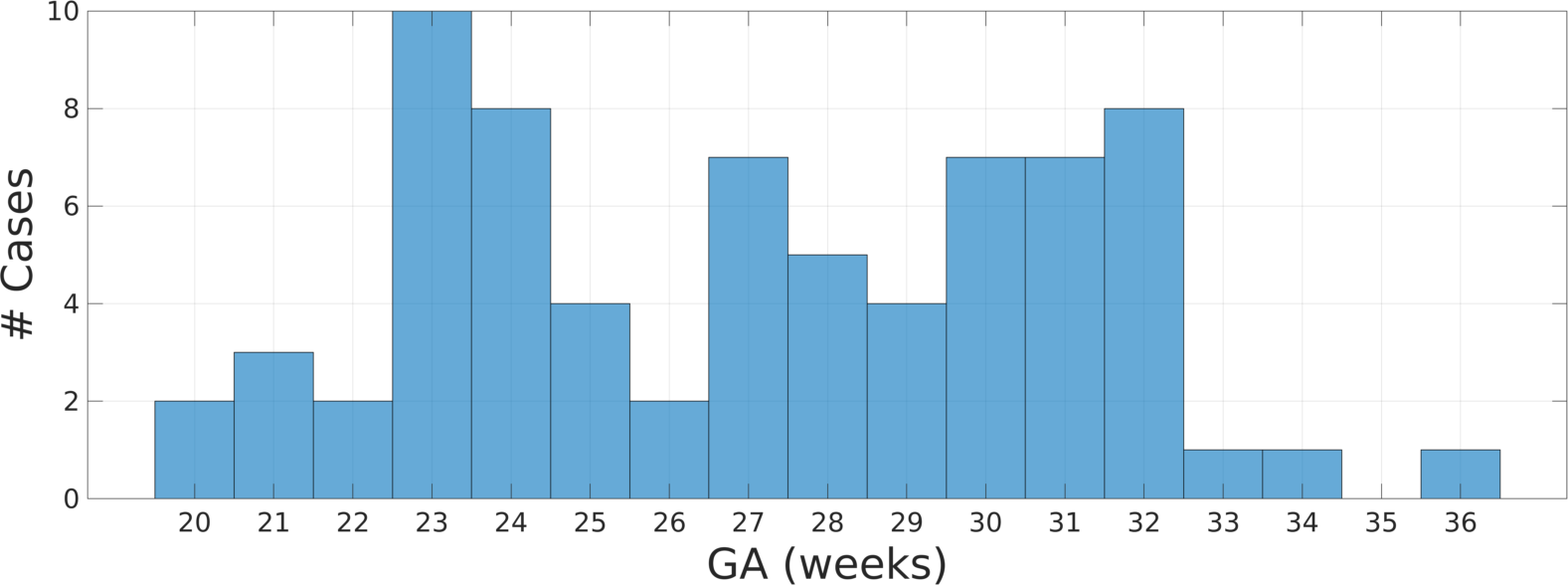}\put(\pamine,\pbmine)
{\makebox(-80,8){\auxminebis\textbf{a)}}}\end{overpic}
\ifdefined\tmiformat
	}\vspace{2mm}\centerline{
\fi
\begin{overpic}[width=0.49\textwidth]{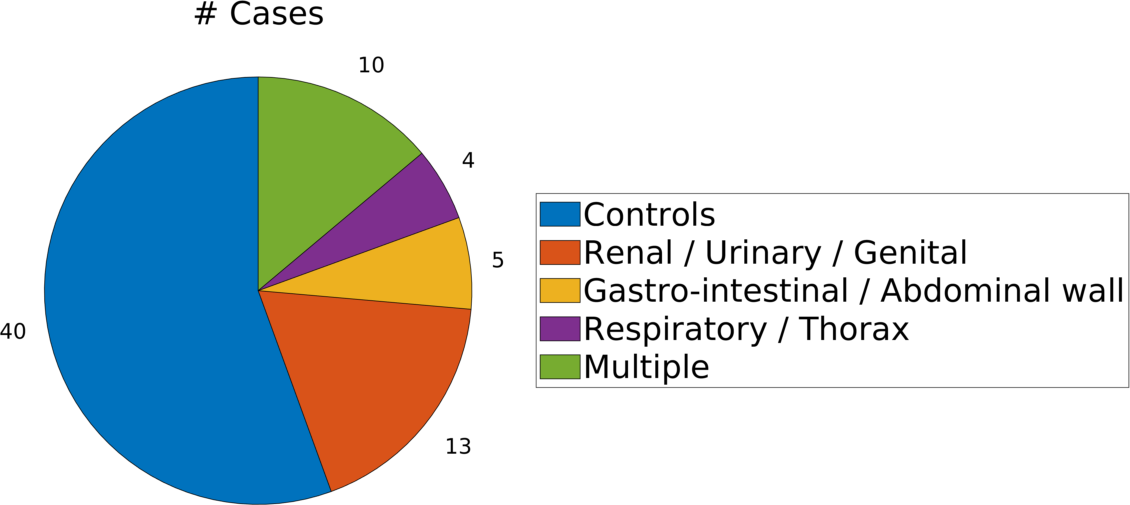}\put(\pamine,\pbmine)
{\makebox(-80,8){\auxminebis\textbf{b)}}}\end{overpic}}
\capsepmine
\caption{Cohort characteristics. \textbf{a)} GA and \textbf{b)} abnormalities distribution.}
\label{fig:FIG07}
\figendmine

Images were acquired on a \textsmaller{\textsc{Philips Ingenia}} $1.5\,\mbox{T}$ with a $24$-channel receive coil using a single-shot fast spin echo sequence. For each subject $N_l=5$ stacks were acquired, including axial (one repeat), sagittal (two repeats) and coronal (two repeats) orientations. Data was collected with an in-plane resolution of $1.25\,\mbox{mm}$ isotropic, $2.5\,\mbox{mm}$ slice thickness and $1.25\,\mbox{mm}$ slice separation. Sensitivity encoding acceleration was set to $2$ with half scan $0.575$ and echo time $T_{\text{E}}=80\,\mbox{ms}$. Number of slices was variable for adequate coverage of the targeted FOV in the range $[100,160]$ with number of packages and interleaves on the order of those in Fig.~\ref{fig:FIG02}a. Total acquisition time was on average $T_{\text{A}}=11'20''$.

\subsection{Implementation details}

\label{sec:IMDE}

We start by performing an OLS reconstruction of the acquired stacks on a $2.5\,\mbox{mm}$ grid. A Region Of Interest (ROI) containing the whole uterus and placenta is drawn on the resulting volumes by the 3D implicit model tool~\citep{Turk02} in the \textsmaller{\textsc{Seg3D}} software~\citep{Seg3D}. This is used to define the FOV of the final reconstruction. Then, the pipeline proceeds as described in Alg.~\ref{alg:ALG01}. At each motion estimation scale $j$ the IRWLS is reset ($\tau=0$) and the algorithm alternates the estimation of the reconstructed image $\mathbf{x}$, DD parameters $\mathbf{\boldsymbol{\theta}}$, prior $\tilde{\mathbf{x}}$, motion $\boldsymbol{\phi}$, and weights $\mathbf{w}$ and $\mathbf{W}$. This is repeated till reaching the target formulation ($\tau=1$) and convergence of motion estimates as dictated by maximum update below a threshold $\delta_{\phi}$. 

The algorithm confronts different subproblems for which a series of parameters need to be selected. First, for robust reconstruction we use a full width half maximum of $2.5\,\mbox{mm}$ for $\boldsymbol{\mathcal{G}}$, appropriate to reduce the variance in data reliability estimation and keep good localization properties, with the simple choice $\kappa=\sqrt{2}$ providing strong suppresion of outliers and acceptable SNR penalty. Second, for registration we use $\beta=2.5$ and $\gamma=2.5$, in agreement with values suggested in the literature~\citep{Faisal05,Zhang19}, $a_{\texttt{STACK}}=0.45$, $a_{\texttt{PACKAGE}}=0.65$ and $a_{\texttt{INTERLEAVE}}=0.85$, which have been observed to prevent early saturation of convergence and guarantee numerical invertibility, and $\sigma_{\text{I}}=0.1$, chosen by visual assessment of plausibility of deformations. Third, for regularization we use $S=2$, achieving a strong suppression of spurious structures while maintaing good data fidelity, visually tuned $\lambda=0.4$, and run $180$ epochs using the Adam optimizer with learning rate $5\cdot 10^{-3}$ for refining the network parameters. Finally, $N_i=5$ iterations are enough for improved IRWLS convergence by homotopy continuation and $\delta_{\phi}$ set to half the in-plane acquisition resolution provides full motion estimation convergence. We refer the reader to the source code for further details.

\begin{algorithm}
\caption{Motion compensated robust reconstruction with deep generative regularization.}
\label{alg:ALG01}
\begin{algorithmic}[1]
\State \textbf{Inputs:} $\mathbf{y}$, $\mathbf{A}$, $\mathbf{P}$, $\boldsymbol{\mathcal{L}}$, $\boldsymbol{\mathcal{G}}$; \textbf{Outputs:} $\mathbf{x}$, $\boldsymbol{\phi}$, $\boldsymbol{\theta}$, $\tilde{\mathbf{x}}$, $\mathbf{w}$
\State $\boldsymbol{\phi}\gets\textbf{0}$, $\mathbf{x}\gets\textbf{0}$, $\tilde{\mathbf{x}}\gets\textbf{0}$, $\boldsymbol{\theta}\gets$ random, $\mathbf{z}\gets$ random
\For{$j\in\{\texttt{STACK,PACKAGE,INTERLEAVE}\}$}
	\State $\tau=0$, $\mathbf{W}\gets\mathbf{I}$
	\While{1}		
		\State $\mathbf{x}\xleftarrow{\eqref{eq:RECP}}$ $\mathbf{W}$, $\tilde{\mathbf{x}}$, $\mathbf{A}$, $\boldsymbol{\phi}^{(j)}$, $\mathbf{x}$, $\mathbf{y}$ (reconstruction)
		\State $\boldsymbol{\theta}\xleftarrow{\eqref{eq:FIWU}}$ $\mathbf{z}$, $\boldsymbol{\theta}$, $\mathbf{x}$ (DD fitting)
		\State $\tilde{\mathbf{x}}\xleftarrow{\eqref{eq:NEOU}}$ $\mathbf{z}$, $\boldsymbol{\theta}$ (DD inference)
		\If{$\tau=1$ and update in $\boldsymbol{\phi}^{(j)}$ lower than $\delta_{\phi}$}
			\State \textbf{break}
		\EndIf
		\State $\boldsymbol{\phi}\xleftarrow{\eqref{eq:DDRR}}$ $\mathbf{P}^{(j)}$, $\boldsymbol{\mathcal{L}}$, $\mathbf{A}$, $\boldsymbol{\phi}^{(j)}$, $\mathbf{x}$, $\mathbf{y}$ (motion estimation)
		\State $\mathbf{w}\xleftarrow{\eqref{eq:WEEQ}}$ $\boldsymbol{\mathcal{G}}$, $\mathbf{A}$, $\boldsymbol{\phi}^{(j)}$, $\mathbf{x}$, $\mathbf{y}$ (weight computation)
		\State $\mathbf{W},\tau\xleftarrow{\eqref{eq:COSH}}\mathbf{w},\tau$ (weight relaxation)
	\EndWhile
	\EndFor
\end{algorithmic}
\end{algorithm}

\subsection{GA prediction}

\label{sec:GAES}

We propose to estimate the GA from the reconstructions by the following steps:
\begin{enumerate}
	\item Obtain a set of slices by uniformly reformatting a reconstructed ROI using $N_r$ rotations evenly distributed in the 3D rotation group. Rotations are performed around the ROI center after windowing the corresponding volumes to prevent boundary artifacts, and three centered slices are extracted along the main reoriented planes.
	\item Extract a set of deep features using a given pre-trained model. We have considered ShuffleNet~\citep{Zhang18}, ResNet-18~\citep{He16}, GoogLeNet~\citep{Szegedy15}, ResNet-50~\citep{He16}, MobileNet-v2~\citep{Sandler18}, ResNet-101~\citep{He16}, and DenseNet-201~\citep{Huang17}, all trained in the ImageNet database~\citep{ImageNet}. Slices extracted in previous step are spatially zero-padded and replicated in the channel dimension to match the input sizes of the models, and z-score normalized.
	\item Taking the deep features as predictors, use zero-correlation constrained linear GA regression~\citep{Treder21}.
	\item At inference time, ensemble the estimates for the $3N_r$ slices into a final GA prediction by taking their median.
\end{enumerate}

\section{Results}

\label{sec:RESU}

In~\S~\ref{sec:VALI} we analyze the contributions of the main constituents of our proposal by an ablation study. In~\S~\ref{sec:CORE} the resolution of our RGDSVR is compared with the state of the art Deformable Slice to Volume Reconstruction (DSVR) method in~\cite{Uus20a}. Clinical utility of both approaches is studied in~\S~\ref{sec:COCU} by presenting results on GA estimation which are also compared with reports from recent approaches.

\subsection{Validation}

\label{sec:VALI}

To assess the impact of the main components of the proposed reconstruction scheme, we compare the reconstructions using the full model ($\mathbf{x}$) versus not using regularization --i.e., $\lambda=0$ in~\eqref{eq:RECP}-- ($\mathbf{x}_{\lambda=0}$); using a handcrafted regularizer $g(\mathbf{x})=\|\boldsymbol{\mathcal{H}}\mathbf{x}\|_2^2$ where $\boldsymbol{\mathcal{H}}$ is the superposition of finite difference penalizers of different orders ($0$ to $32$) whose weights have been visually tuned for trading off noise suppression and resolution loss ($\mathbf{x}_{\mathcal{H}}$); not using the robust formulation --i.e., fixing $\tau=0$ in~\eqref{eq:COSH}-- ($\mathbf{x}_{\tau=0}$); and not using motion correction --i.e., $\sigma_{\text{I}}\to\infty$ in~(\ref{eq:GRAM})-- ($\mathbf{x}_{\sigma_{\text{I}}\to\infty}$). The convergence criterion has been adjusted for comparable number of iterations for all alternatives. We report results for the scan with median level of data corruption as assessed by the normalized cross correlation of adjacent input slices, an indicator that has been reported to agree with human observer quality ratings~\citep{Uus20a}. Results for other subjects are included as Supporting Information.

We observe that reconstructions without regularization (Fig.~\ref{fig:FIG04}b) present a noisy appearance. Noise can be mitigated either by the regularizer penalizing high-frequency content (Fig.~\ref{fig:FIG04}c) or by our proposed DD regularizer in Fig.~\ref{fig:FIG04}a, but resolution is better preserved when using the DD regularizer, for instance at the boundaries between the fetal body and the amniotic fluid in the area enclosed in blue. Ability to resolve fine detailed structures is evident when comparing to results without motion compensation (Fig.~\ref{fig:FIG04}e). Non-robust reconstructions (Fig.~\ref{fig:FIG04}d) look similar to robust reconstructions in Fig.~\ref{fig:FIG04}a. However, impact of non-suppressed artifacts is noticeable locally, as in the area within the blue ellipse, where more uniform fluid background is observed when the robust formulation is adopted.

\figstaastmine
\renewcommand{\pamine}{50}
\ifdefined\tmiformat
	\setlength{\hcolw}{0.19\textwidth}
\else
	\setlength{\hcolw}{0.17\textwidth}
\fi	
\centerline{
\begin{overpic}[width=\hcolw]{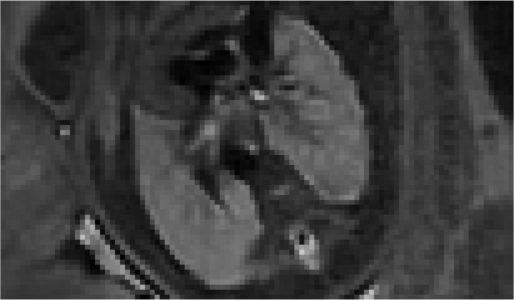}\end{overpic}
\begin{overpic}[width=\hcolw]{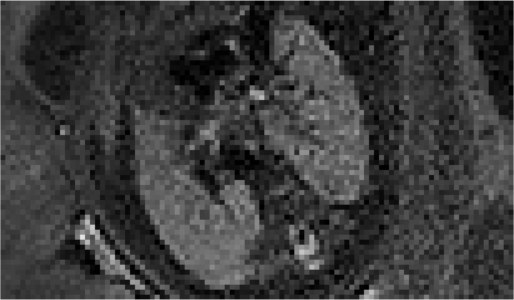}\end{overpic}
\begin{overpic}[width=\hcolw]{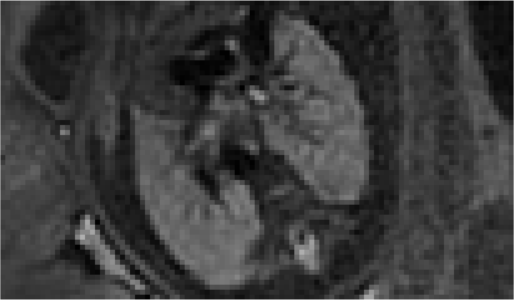}\end{overpic}
\begin{overpic}[width=\hcolw]{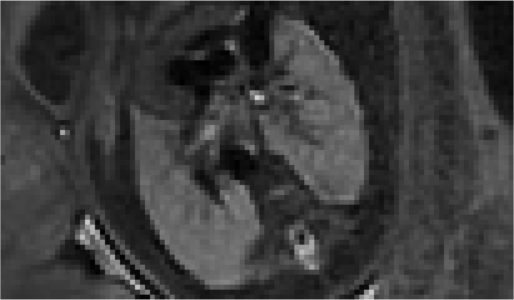}\end{overpic}
\begin{overpic}[width=\hcolw]{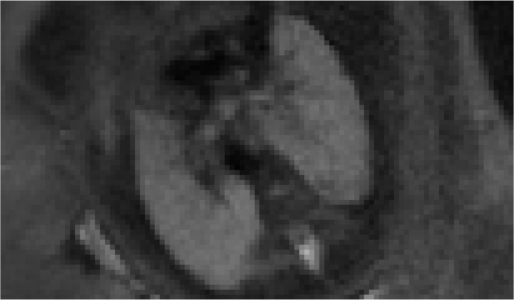}\end{overpic}\\}
\vspace{1mm}
\centerline{
\begin{overpic}[width=\hcolw]{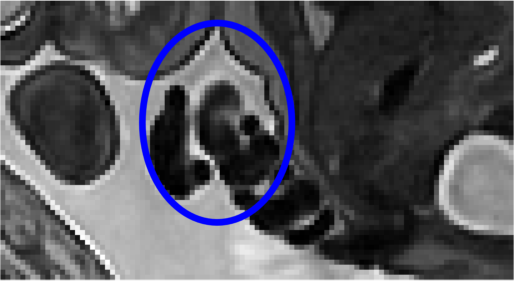}\end{overpic}
\begin{overpic}[width=\hcolw]{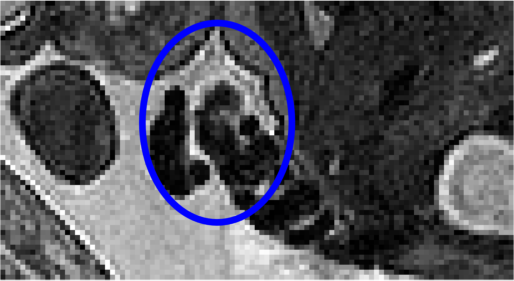}\end{overpic}
\begin{overpic}[width=\hcolw]{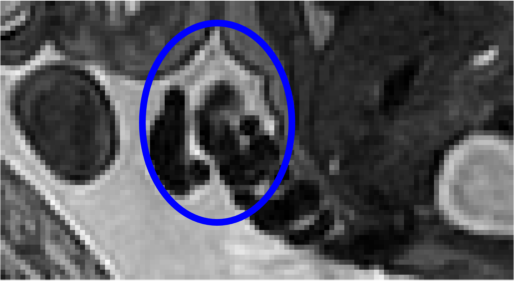}\end{overpic}
\begin{overpic}[width=\hcolw]{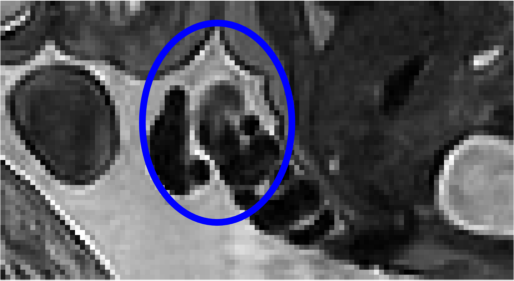}\end{overpic}
\begin{overpic}[width=\hcolw]{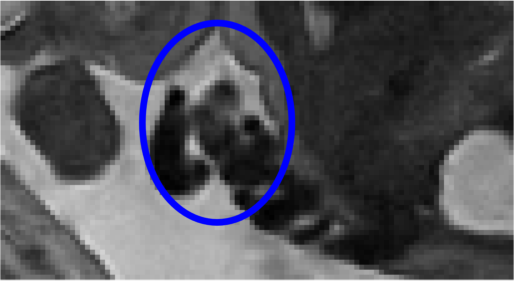}\end{overpic}\\}
\vspace{1mm}
\centerline{
\begin{overpic}[width=\hcolw]{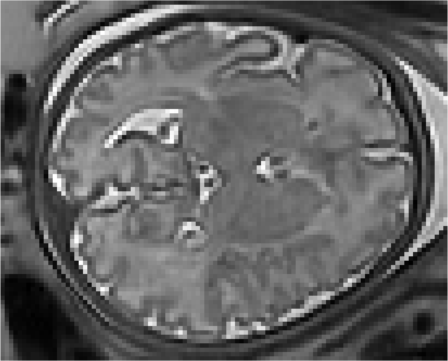}\put(\pamine,\pbmine)
{\makebox(-90,-2){\auxminebis\textbf{a)}}}\end{overpic}
\begin{overpic}[width=\hcolw]{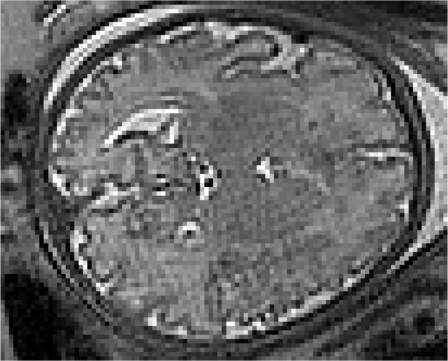}\put(\pamine,\pbmine)
{\makebox(-90,-2){\auxminebis\textbf{b)}}}\end{overpic}
\begin{overpic}[width=\hcolw]{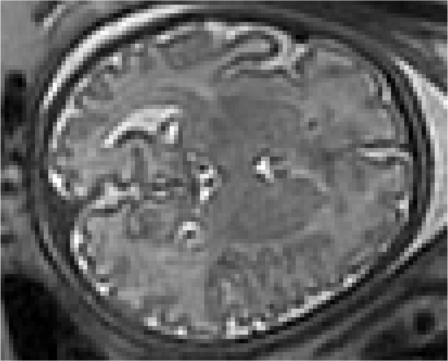}\put(\pamine,\pbmine)
{\makebox(-90,-2){\auxminebis\textbf{c)}}}\end{overpic}
\begin{overpic}[width=\hcolw]{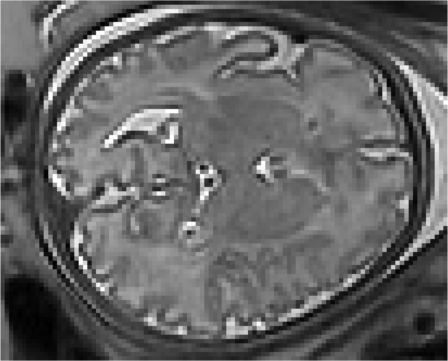}\put(\pamine,\pbmine)
{\makebox(-90,-2){\auxminebis\textbf{d)}}}\end{overpic}
\begin{overpic}[width=\hcolw]{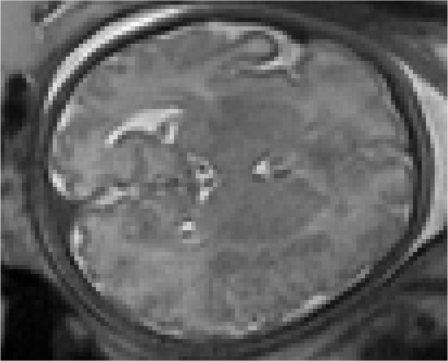}\put(\pamine,\pbmine)
{\makebox(-90,-2){\auxminebis\textbf{e)}}}\end{overpic}\\
}
\capsepmine
\caption{Comparison of different reconstruction alternatives in the case with median degradation. From top to bottom, coronal, sagittal and axial planes in the mother's geometry. Reconstructions \textbf{(a)} based on the full model ($\mathbf{x}$); \textbf{(b)} without regularization ($\mathbf{x}_{\lambda=0}$); \textbf{(c)} with handcrafted regularization ($\mathbf{x}_{\mathcal{H}}$); \textbf{(d)} without the robust formulation ($\mathbf{x}_{\tau=0}$); \textbf{(e)} without motion correction ($\mathbf{x}_{\sigma_{\text{I}}\to\infty}$). The blue ellipse highlights local artifacts, noise or blurring around the umbilical chord when taking off any of the main components of our formulation.}
\label{fig:FIG04}
\figendastmine

In Fig.~\ref{fig:FIG05}a we show the power spectral density (PSD) averaged along the three axes of the reconstruction grid for the aforementioned reconstruction alternatives applied to the same subject. Reconstructions without regularization ($\mathbf{x}_{\lambda=0}$) present the largest power at high spatial frequencies with disruption of power law of attenuation above approximately $1.5\,\mbox{mm}$, probably stemming from a Gibbs ringing filter applied as part of the scanner k-space reconstructions. However, we know from image inspection in Fig.~\ref{fig:FIG04}b that a significant amount of the energy at high frequencies is contributed by noise. Noise reduction was effective when using the handcrafted regularizer, but we observe here ($\mathbf{x}_{\mathcal{H}}$) that this comes at the price of strong suppression of high frequency signal components. The DD-based regularized reconstructions, denoted by $\mathbf{x}$, lie somewhere in between both scenarios, likely with better signal preservation at high frequencies and strong noise reduction. Small differences are observed between non-robust ($\mathbf{x}_{\tau=0}$) and robust reconstructions, but the non-robust version presents slower PSD decay rates consistent with reduced suppression of artifactual structures. Motion compensation has an impact in moderate to high spatial frequencies, with power enhancements versus non-compensated reconstructions ($\mathbf{x}_{\sigma_{\text{I}}\to\infty}$) above $5\,\mbox{dBs}$ at $2\,\mbox{mm}$, for instance.

\figstamine
\renewcommand{\pamine}{50}
\vspace{2mm}
\centerline{
\begin{overpic}[width=0.49\textwidth]{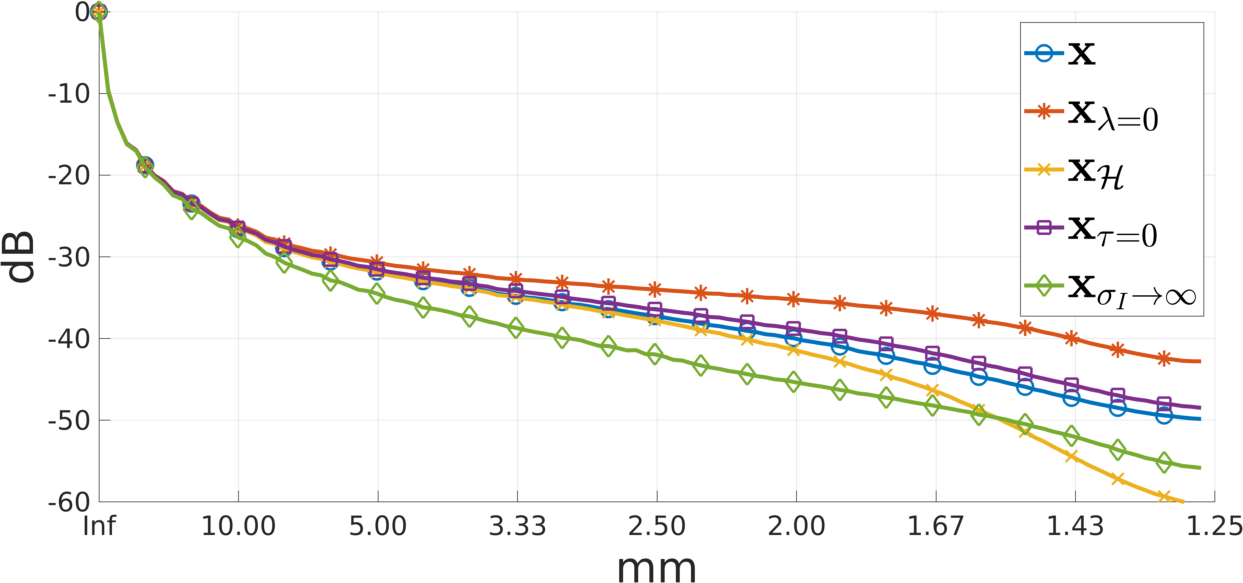}\put(\pamine,\pbmine)
{\makebox(-75,8){\auxminebis\textbf{a)}}}\end{overpic}
\ifdefined\tmiformat
	}\vspace{2mm}\centerline{
\fi
\begin{overpic}[width=0.49\textwidth]{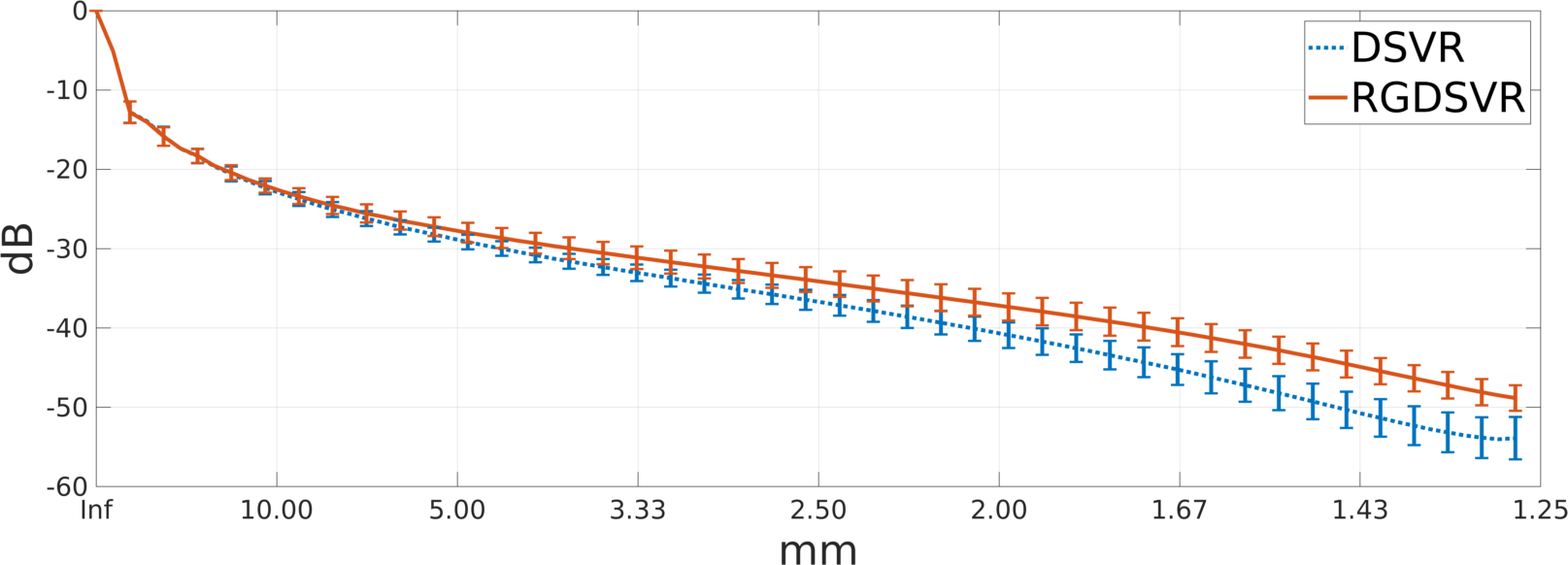}\put(\pamine,\pbmine)
{\makebox(-75,8){\auxminebis\textbf{b)}}}\end{overpic}}
\capsepmine
\caption{PSD comparisons. \textbf{a)} Reconstruction alternatives in the case with median degradation (uterus ROI); \textbf{b)} DSVR and RGDSVR (trunk ROI).}
\label{fig:FIG05}
\figendmine

\subsection{Comparison with the literature: resolution}

\label{sec:CORE}

We compare the resolution provided by our RGDSVR and the DSVR method in~\cite{Uus20a}. The reconstruction grid orientation can be different for both methods, so we used sinc interpolation to reformat our datasets into the grid of the reference method. In Fig.~\ref{fig:FIG05}b we show the mean$\pm$std PSD of both methods across the cohort described in~\S~\ref{sec:MATE}. The curves suggest better preserved moderate to high spatial frequency information when using RGDSVR, with significant differences observable for structures below $15\,\mbox{mm}$ as confirmed by paired right-tailed sign tests against the null hypothesis that the median of the difference between the PSD of the RGDSVR and DSVR is lower than zero or zero ($p<0.05$), and highly significant for structures below $10\,\mbox{mm}$ ($p\ll 0.05$).

In Fig.~\ref{fig:FIG06} we visually compare the results of both methods. Despite reconstructions are not spatially matched due to arbitrary definitions of reference deformations, we observe that RGDSVR tends to provide sharper results for comparable levels of noise and artifacts. This is illustrated by the ellipses on top of the main vessels, the bowel, and the neck, respectively in the 
\ifdefined\tmiformat
top, central and bottom
\else
left, central and right
\fi
 panels.

\figstamine
\renewcommand{\pamine}{50}
\setlength{\hcolw}{0.24\textwidth}
\ifdefined\tmiformat
	\centerline{
	\begin{overpic}[width=\hcolw,trim={0cm 3cm 0cm 2.5cm},clip]{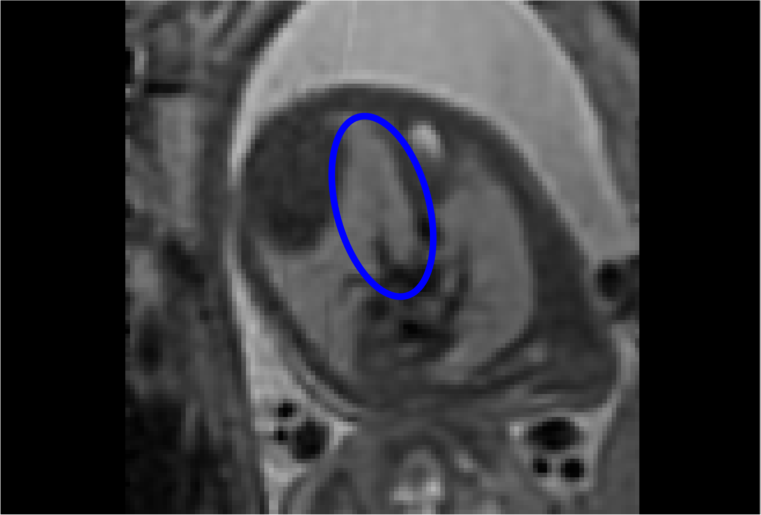}\end{overpic}
	\begin{overpic}[width=\hcolw,trim={0cm 3cm 0cm 2.5cm},clip]{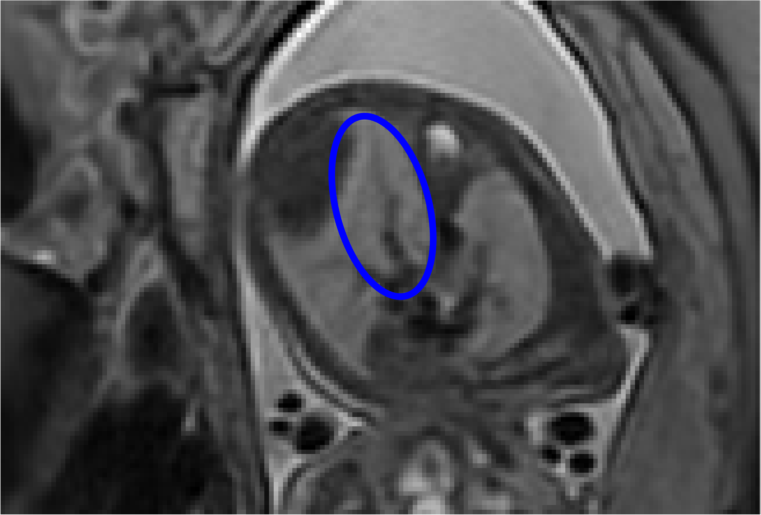}\end{overpic}\\}
	\vspace{1mm}
	\centerline{
	\begin{overpic}[width=\hcolw,trim={0cm 3cm 0cm 2cm},clip]{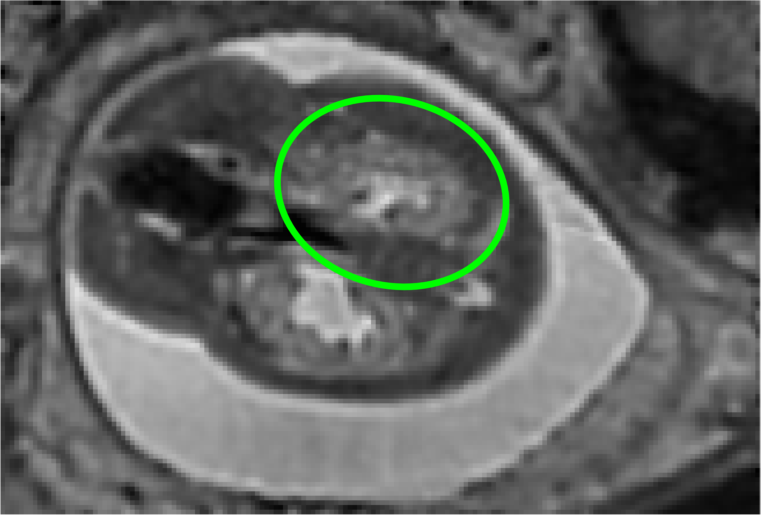}\end{overpic}
	\begin{overpic}[width=\hcolw,trim={0cm 3cm 0cm 2cm},clip]{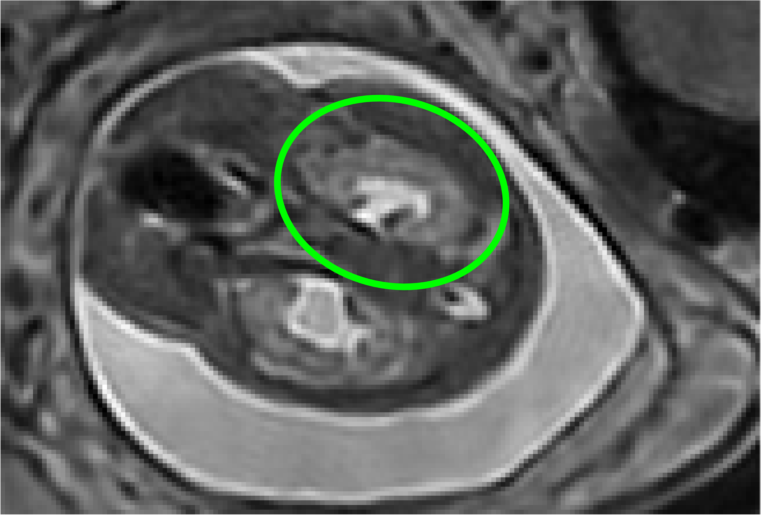}\end{overpic}\\}
	\vspace{1mm}
	\centerline{
	\begin{overpic}[width=\hcolw,trim={0cm 0cm 0cm 2cm},clip]{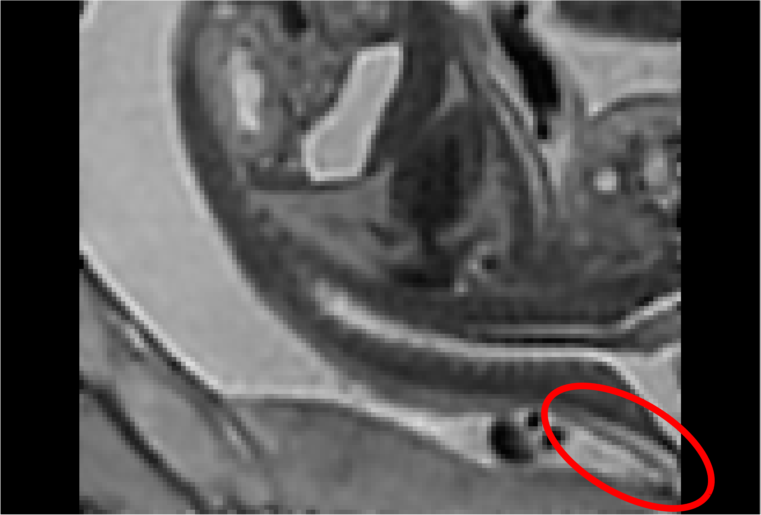}\put(\pamine,\pbmine)
	{\makebox(-90,0){\auxminebis\textbf{a)}}}\end{overpic}
	\begin{overpic}[width=\hcolw,trim={0cm 0cm 0cm 2cm},clip]{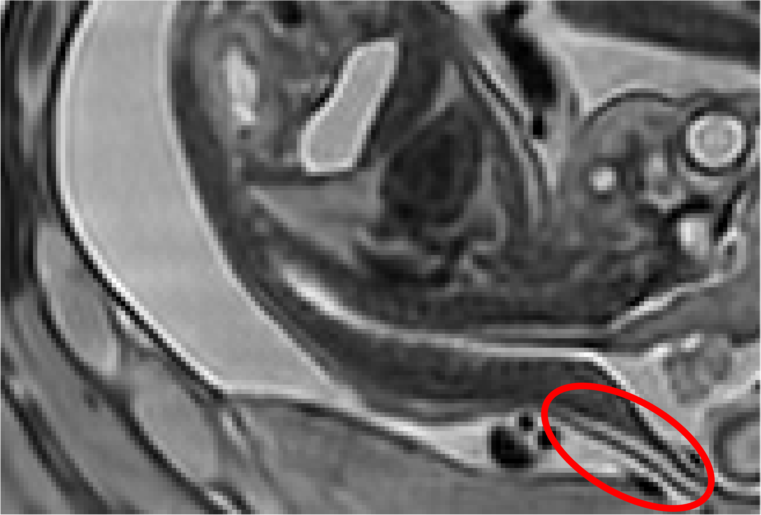}\put(\pamine,\pbmine)
	{\makebox(-90,0){\auxminebis\textbf{b)}}}\end{overpic}}
\else
	\centerline{
	\begin{overpic}[width=\hcolw,trim={0cm 3cm 0cm 2.5cm},clip]{figs/fig06/fig06-1-1}\end{overpic}
	\begin{overpic}[width=\hcolw,trim={0cm 3cm 0cm 2cm},clip]{figs/fig06/fig06-1-2}\end{overpic}
	\begin{overpic}[width=\hcolw,trim={0cm 0cm 0cm 2cm},clip]{figs/fig06/fig06-1-3}\put(\pamine,\pbmine)
	{\makebox(-500,0){\auxminebis\textbf{a)}}}\end{overpic}\\}
	\vspace{1mm}
	\centerline{
	\begin{overpic}[width=\hcolw,trim={0cm 3cm 0cm 2.5cm},clip]{figs/fig06/fig06-2-1}\end{overpic}
	\begin{overpic}[width=\hcolw,trim={0cm 3cm 0cm 2cm},clip]{figs/fig06/fig06-2-2}\end{overpic}
	\begin{overpic}[width=\hcolw,trim={0cm 0cm 0cm 2cm},clip]{figs/fig06/fig06-2-3}\put(\pamine,\pbmine)
	{\makebox(-500,0){\auxminebis\textbf{b)}}}\end{overpic}\\}
\fi
\capsepmine
\caption{Three orthogonal views with reconstructions using \textbf{a)} DSVR and \textbf{b)} RGDSVR. Areas enclosed by the  
\ifdefined\tmiformat
	blue (top panel), green (central panel) and red (bottom panel)
\else 
	blue (left panel), green (central panel) and red (right panel)
\fi 
 ellipses suggest that our method provides improved conspicuity of the main vessels, sharper contrast and better delineation of the bowel loops, and increased resolution of stacked tissue layers in the neck, respectively.}
\label{fig:FIG06}
\figendmine

\subsection{Clinical application: GA prediction}

\label{sec:COCU}

Despite potential gains of proposed reconstructions are suggested by the resolution comparison in~\S~\ref{sec:CORE}, lack of ground truth makes direct comparison of methods extremely challenging. In addition, it is very difficult to faithfully reproduce the different sources of corruption in real data as well as complex patterns of fetal and mother motion by simulations. Therefore, we have resorted to a task-based validation where we perform comparisons on a clinically-oriented GA prediction problem. We compare the performance of RGDSVR and DSVR for GA estimation using a trunk ROI corresponding to the FOV returned by the default DSVR implementation. As existing methods for fetal GA estimation from MRI~\citep{Liao20,Shi20} are based on brain data, we also test the GA estimation performance using a brain ROI from our reconstructions. In all cases we use the GA estimation method described in~\S~\ref{sec:GAES} with 3D space spanned by slices from $N_r=200$ random volume reorientations. Finally, to investigate the added value of non-brain features, we combine brain and trunk ROIs by using $N_r=100$ random reorientations each. In this case, common regression and z-score normalization weights are computed at training using slices from both ROIs, and joint median ensemble of estimates is used at inference.

We perform a 6-fold cross validation using the $72$ cases in our cohort. Different GA regression alternatives are compared by computing the Mean Absolute Error (MAE) and coefficient of determination $R^2$ with results reported in Table~\ref{tab:TAB01}. GA predictions using RGDSVR consistently outperform predictions using DSVR for comparable trunk ROI. Best results in both cases are obtained using DenseNet-201, respectively with $\mbox{MAE}$ ($R^2$) $0.931$ ($0.918$) and $1.045$ ($0.888$). We also observe consistently better results for all models when using the brain rather than the trunk ROI, with best figures $0.683$ ($0.950$) provided again by DenseNet-201. These results compare favourably with those reported by~\cite{Liao20}, $0.751$ ($0.947$), and~\cite{Shi20}, $0.767$ ($0.920$), and, in terms of MAE, they seem to do so by a substantial margin. Further to this, with the exception of $R^2$ for poorly performing MobileNet-v2 and GoogleLeNet, additional improvements are consistently observed for all models when combining volumetric brain and trunk information, a unique feature of the proposed technique, with $\mbox{MAE}$ ($R^2$) $0.618$ ($0.958$) for DenseNet-201. Ground truth GA is obtained from the clinical records which is not free from errors. Therefore, in Fig.~\ref{fig:FIG08} we provide Bland-Altman plots of agreement~\citep{Bland99} between DenseNet-201 based predictions ($\text{GA}_{\text{DN}}$) and ground truth GA ($\text{GA}_{\text{GT}}$) for the considered reconstructions and ROIs combinations. Results are color coded according to the abnormality categories legend in Fig.~\ref{fig:FIG07}b. We observed no statistical significance at $p=0.05$ in GA discrepancies between both methods when comparing controls and cases with anomalies with an unequal variances $t$-test for any of the alternatives considered in Fig.~\ref{fig:FIG08}.

\tabstaastmine
\begin{center}
\begin{tabular}{c||c|c||c|c||c|c||c|c}
 & \multicolumn{2}{c||}{\shortstack{DSVR\\trunk}} & \multicolumn{2}{c||}{\shortstack{RGDSVR\\trunk}} & \multicolumn{2}{c||}{\shortstack{RGDSVR\\brain}} & \multicolumn{2}{c}{\shortstack{RGDSVR\\brain \& trunk}} \\\hline
 & $\mbox{MAE}$ & $R^2$ & $\mbox{MAE}$ & $R^2$ & $\mbox{MAE}$ & $R^2$ & $\mbox{MAE}$ & $R^2$ \\\hline\hline
ShuffleNet & $1.115$ & $0.874$ & $1.012$ & $0.895$ & $0.764$ & $0.941$ & $0.632$ & $0.956$ \\\hline
ResNet-18 & $1.234$ & $0.829$ & $1.133$ & $0.876$ & $0.712$ & $0.941$ & $0.674$ & $0.947$ \\\hline
GoogLeNet & $1.472$ & $0.771$ & $1.049$ & $0.885$ & $0.714$ & $0.945$ & $0.694$ & $0.945$ \\\hline
ResNet-50 & $1.164$ & $0.841$ & $0.955$ & $0.910$ & $0.734$ & $0.937$ & $0.686$ & $0.948$ \\\hline
MobileNet-v2 & $1.201$ & $0.841$ & $1.137$ & $0.863$ & $0.712$ & $0.945$ & $0.705$ & $0.943$ \\\hline
ResNet-101 & $1.218$ & $0.839$ & $1.015$ & $0.900$ & $0.731$ & $0.943$ & $0.677$ & $0.949$ \\\hline
DenseNet-201 & $1.045$ & $0.888$ & $0.931$ & $0.918$ & $0.683$ & $0.950$ & $\mathbf{0.618}$ & $\mathbf{0.958}$ 
\end{tabular}
\end{center}
\caption{MAE (weeks) and $R^2$ in GA estimation using different reconstructions, ROIs, and deep features. Best results are boldfaced.}
\label{tab:TAB01}
\tabendastmine

\figstaastmine
\renewcommand{\pamine}{50}
\setlength{\hcolw}{0.244\textwidth}
\centerline{
\begin{overpic}[width=\hcolw]{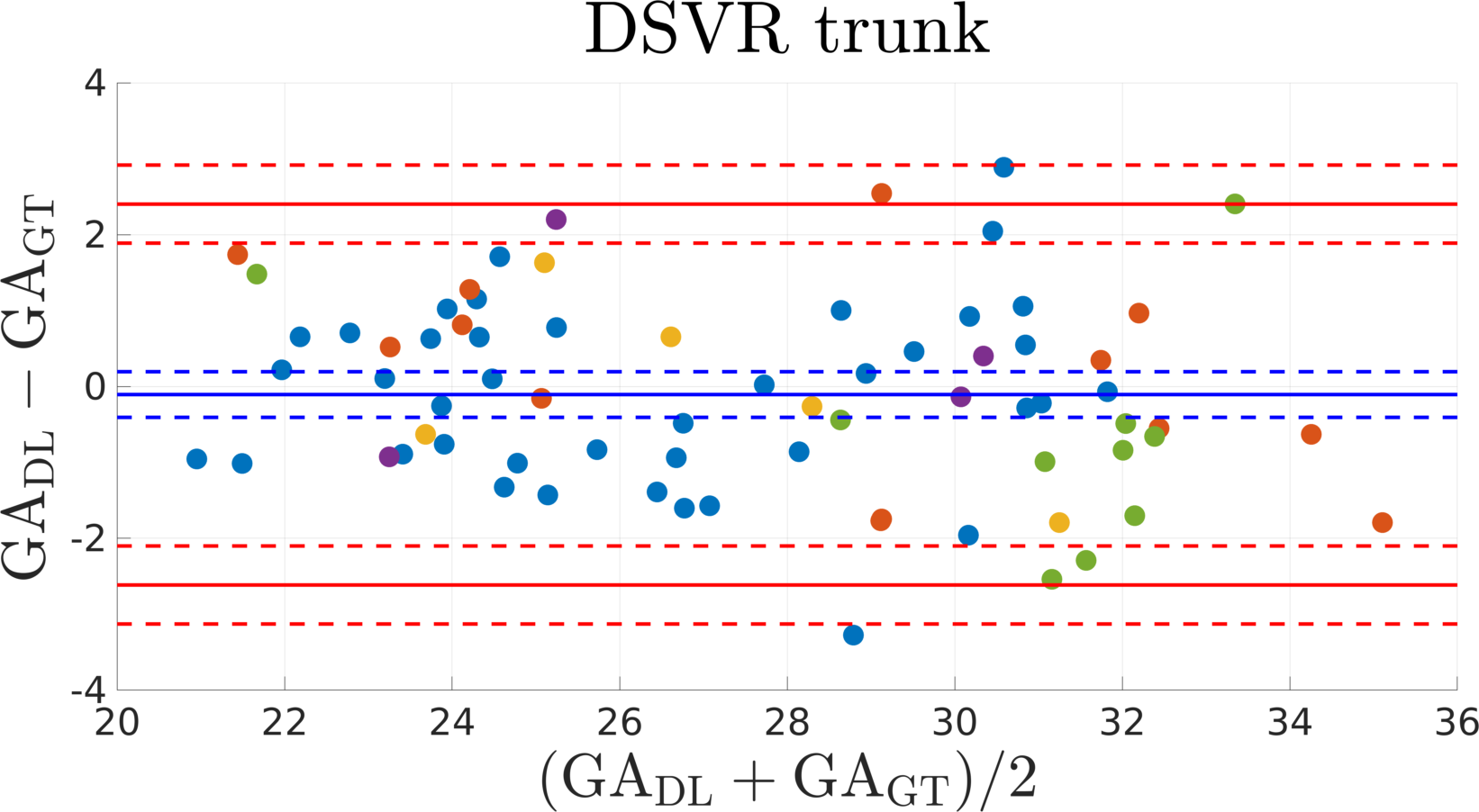}\put(\pamine,\pbmine)
{\makebox(-80,10){\auxminebis\textbf{a)}}}\end{overpic}
\begin{overpic}[width=\hcolw]{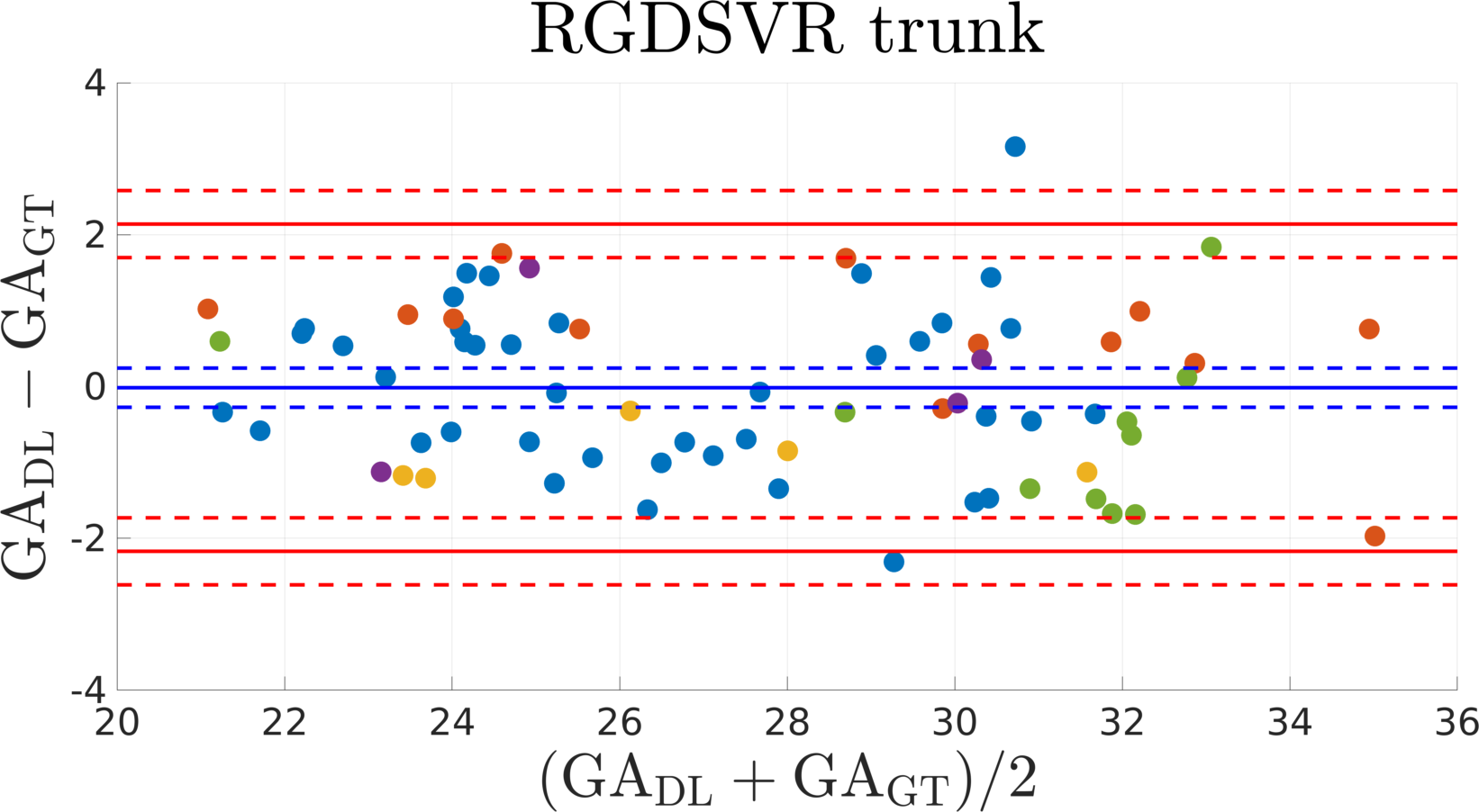}\put(\pamine,\pbmine)
{\makebox(-80,10){\auxminebis\textbf{b)}}}\end{overpic}
\begin{overpic}[width=\hcolw]{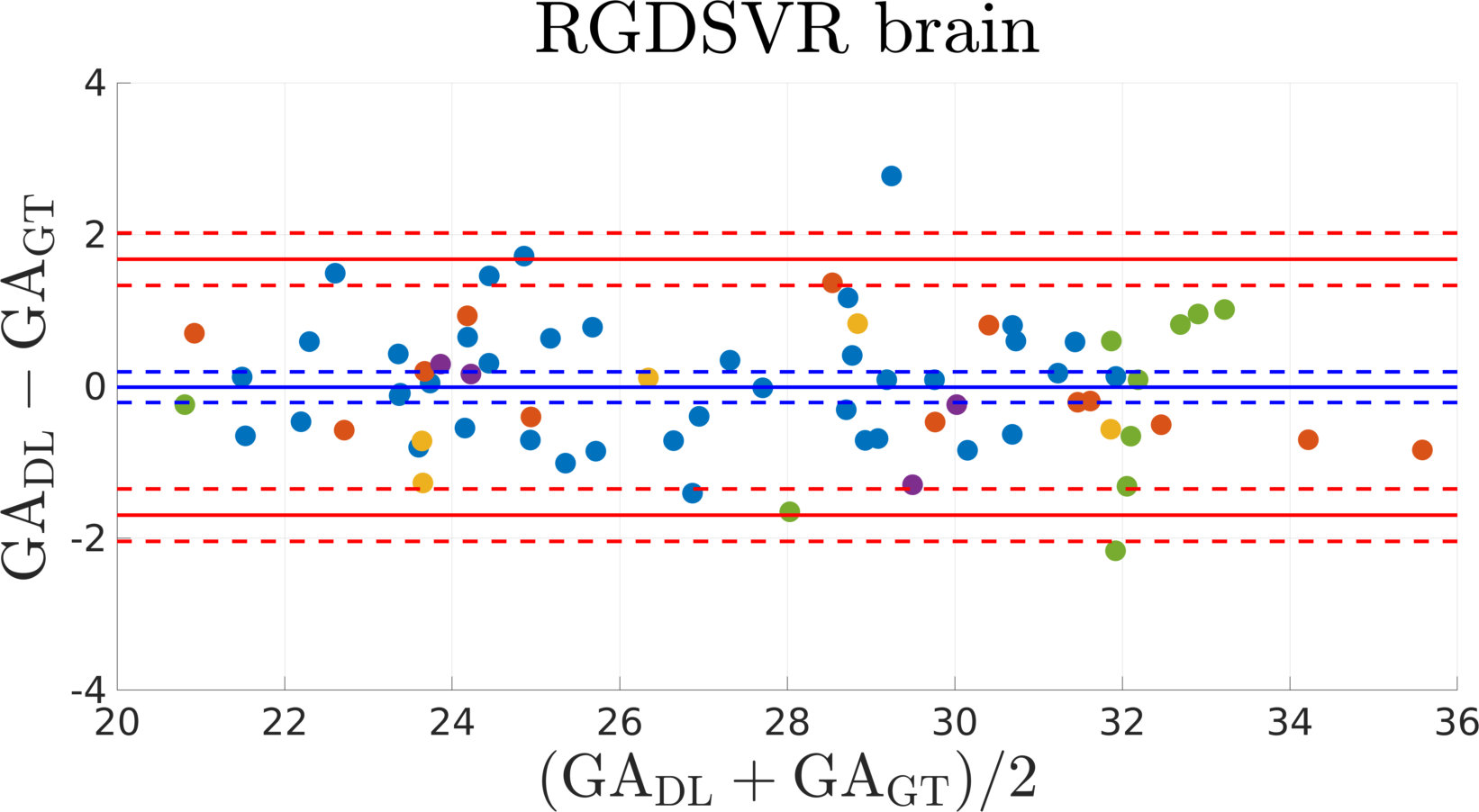}\put(\pamine,\pbmine)
{\makebox(-80,10){\auxminebis\textbf{c)}}}\end{overpic}
\begin{overpic}[width=\hcolw]{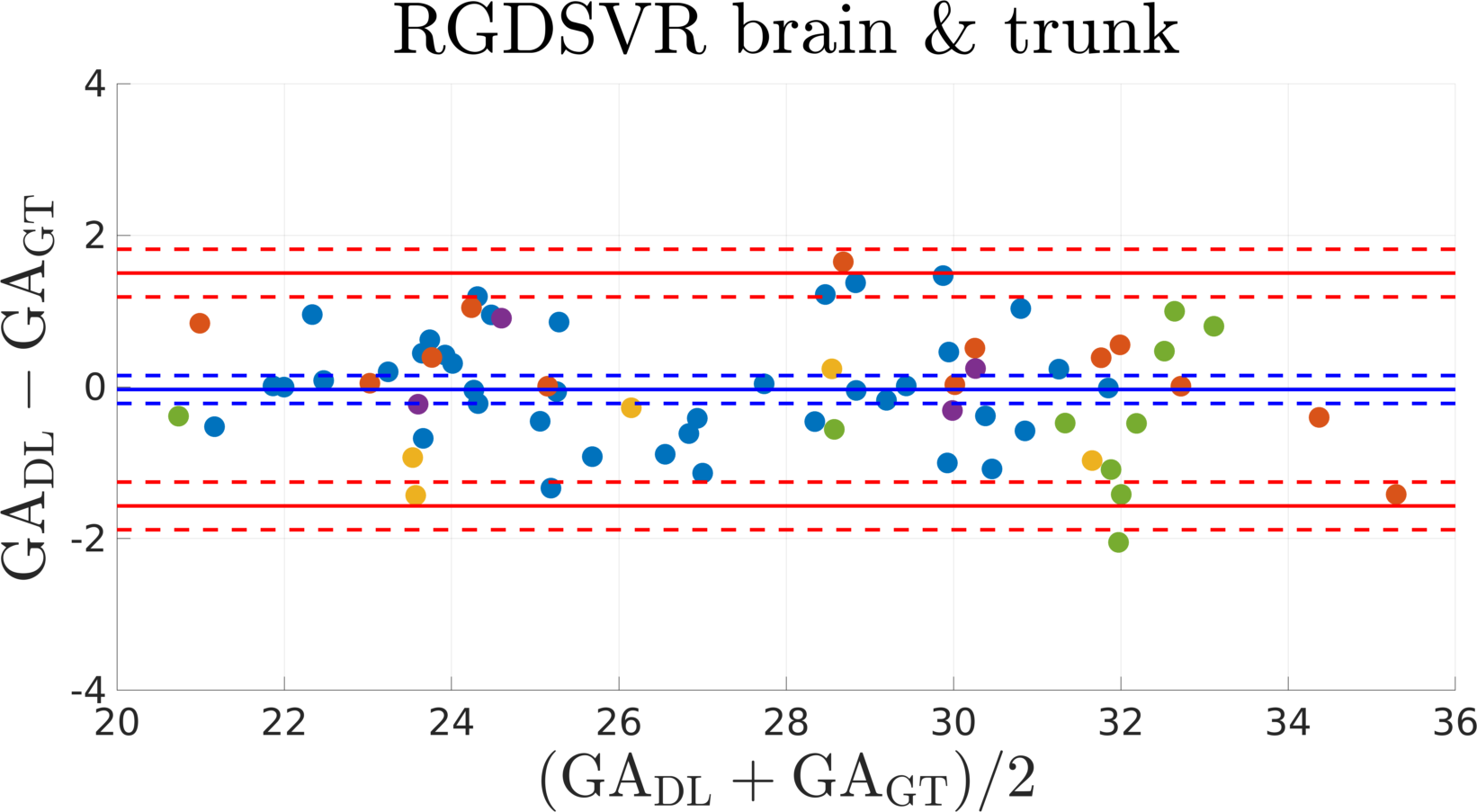}\put(\pamine,\pbmine)
{\makebox(-80,10){\auxminebis\textbf{d)}}}\end{overpic}
}
\caption{Bland-Altman plots for DenseNet-201. \textbf{a)} DSVR trunk.  \textbf{b)} RGDSVR trunk.  \textbf{c)} RGDSVR brain.  \textbf{d)} RGDSVR brain \& trunk. Solid blue line: mean differences. Solid red lines: $95\%$ limits of agreement. Dashed lines: corresponding approximate confidence intervals~\citep{Bland99}.}
\label{fig:FIG08}
\figendastmine

\section{Discussion}

\label{sec:DISC}

We have proposed a novel methodology for robust whole-body fetal MRI reconstruction relying on diffeomorphic motion estimation to capture plausible deformations of the fetal organs and high-quality regularization using a deep generative model. We have quantitatively and qualitatively characterized the impact of these components in the reconstructions. Comparisons with a state of the art method have demonstrated noticeable differences in reconstruction sharpness and suggested a strong impact of the reconstruction method in the clinical utility of fetal MRI, which has been showcased by a GA prediction task. We have provided a conceptually simple GA prediction method based on free reformatting the 3D reconstructions for 2D deep feature extraction and correlation constrained linear regression, showing improved accuracy with respect to existing approaches.

Similar to~\cite{Gholipour10}, we have built our cost function using the robust regression via M-estimators framework, so weights for outlier mitigation are directly derived from the cost function~\citep{Maronna19}, instead of using independent metrics as in~\cite{Uus20a}. In addition, the introduction of the smoothing operator $\boldsymbol{\mathcal{G}}$ simplifies previous combinations of voxelwise and slicewise weights~\citep{Gholipour10,Uus20a} by assuming that motion-induced degradation of magnetization as well as uncorrected non-rigid motion have a regional nature.

In the non-rigid registration setting, reproducible morphometry may be compromised by geometric distortions introduced by the algorithm, which can be alleviated by the diffeomorphic constraint. Of particular interest, we should highlight the brain; despite the prominent use of rigid motion models in the past, non-rigid components may become appropriate to model non-linearities of the scanner gradient fields~\citep{Doran05}. Although our motion model should probably be refined for the brain, for instance via decoupled motion models~\cite{Uus20b}, distortion levels in the reconstructions are small enough so as to lead to accurate GA estimates.

Taking into account the limitations for generating ground-truth datasets, we have opted for an unsupervised application of deep architectures for regularizing our reconstructions. However, there are alternatives for integration of DL into reconstruction problems~\cite{Ongie20} and DL could also be applied for extending the capture range of registration or refining the characterization of outliers. On the other hand, the memory footprint of the 3D DD architecture is a strong limitation, so deep network regularizers based on implicit representations~\citep{Fathony21} could be considered.

3D reconstruction errors in whole-body fetal MRI may arise from multiple causes. We may encounter errors due to (a) small inaccuracies in motion estimation, (b) inconsistencies in magnetization of different slices, (c) multiple poses of the fetal body throughout the scan, and (d) the fetus moving continuously across the examination. We believe that in cases (a) and (b) information coming from complementary stacks can resolve the ambiguities in many instances, so the robust formulation and deep generative regularization generally provides satisfactory solutions. However, artifacts in the reconstructions may be strong in cases (c) and (d), as the method may struggle to find a good direction for high quality convergence.

Reconstruction quality is ultimately determined by available scan time, as enlarged sampling redundancy gives more flexibility to implement robust reconstructions with improved SNR. For a fixed total acquisition time, there are different acquisition choices that may impact the reliability of fetal reconstructions. Importantly,~\cite{Shilling08} studies the comparative performance of overlapped single orientation scans versus multi-oriented scans in terms of resolution retrieval, with clear benefits observed in the latter (see also~\citep{Reeth12}). In our context, this study may suggest replacing the repeated sagittal and coronal stacks by new orientations, perhaps whilst changing the overlapping factors. However, implementation of theoretically optimal sampling schemes is often limited by hardware specifications of the scanner, inherent complexity of fetal imaging in vivo, computational requirements of reconstruction algorithms, or need of harmonization with protocols currently in place.

Our GA estimation method leverages free reformatting of volumetric reconstructions to obtain a dense set of slices covering the fetal structures in the variable spatial configurations they can adopt due to fetal motion. We have shown that deep feature extraction using pre-trained models combined with correlation constrained linear regression provides accurate results for this task. Our results look superior to existing methods~\citep{Liao20,Shi20}, particularly when complementing brain features with trunk information, but there are differences in the cohorts considered. Most notably, existing methods use larger cohorts including single-sequence data from $289$ subjects~\citep{Liao20} and multi-sequence data from $764$ subjects~\citep{Shi20}. Small sample sizes in our case precluded isolation of a subset of subjects for testing, a potentially important limitation when compared to~\cite{Shi20}. Finally, our GA estimation pipeline permits a straightforward application to brain age prediction in adults~\citep{Cole17}, particularly if volumetrically encoded sequences are available.

In the future we plan to follow a practical roadmap to facilitate the application of our algorithm in clinical scenarios. This may include automated ROI extraction, refining and further testing the algorithm and GA estimation using additional cohorts and acquisition protocols, and move towards a comprehensive analysis pipeline by integrating techniques for whole-body fetal segmentation and atlas construction~\citep{Torrents-Barrena21}.

\section{Conclusions}

\label{sec:CONC}

We have proposed a method for robust whole-body fetal and placenta MRI based on diffeomorphic registration and deep generative regularization. Volumetric reconstructions are obtained from a set of motion-affected and possibly corrupted single-shot slices. Our proposal provides alternative solutions to existing methods for the different subproblems faced in this application. These are validated by an ablation study and improved conspicuity is shown when compared with a state of the art method in a cohort of $72$ fetal subjects. A GA estimation task is defined to assess the clinical utility of our technique, for which we propose a simple method leveraging the 3D information, which produced competitive results. For usual levels of motion, our reconstructions provide dense and consistent representations of the fetal anatomy. Therefore, the proposed methods may find application in 3D fetal MRI morphometry, developmental assessment, or fetal surgery planning.

\ifdefined\tmiformat
	\bibliographystyle{IEEEtran}
\else	
	\section*{Acknowledgments}

	This work is funded by the Ministry of Science and Innovation, Spain, under the Beatriz Galindo Programme [BGP18/00178]. This work has been supported by the Madrid Government (Comunidad de Madrid-Spain) under the Multiannual Agreement with Universidad Polit\'ecnica de Madrid in the line Support for R\&D projects for Beatriz Galindo researchers, in the context of the V PRICIT (Regional Programme of Research and Technological Innovation).
	
	\clearpage
	\bibliographystyle{apalike}
\fi

\bibliography{RGDSVR}

\clearpage

\ifdefined\tmiformat
	\appendices
	\pagestyle{fancy}
	\pagenumbering{Roman} 
	\setcounter{page}{1}

	\renewcommand\thefigure{\Roman{figure}}
	\setcounter{figure}{0}
	
	\fancyhf{}
	\fancyhead[LE,RO]{}
	\fancyhead[RO,LE]{\thepage}
\else
	\appendix
\fi

\section*{Supplementary material}

\label{sec:SUMA}

In~\S~\ref{sec:VALI} of the main manuscript we presented the results of an ablation study using the scan with median level of data corruption as assessed by the normalized cross correlation of adjacent input slices~\citep{Uus20a}. Here we show corresponding results for exemplary scans at different levels of collected slices corruption. Namely, we present results for scans with lowest, $25$-percentile, $75$-percentile, and largest input corruption levels.

In Fig.~\ref{fig:FIGI} we show the original stacks in mid-planes along the readout direction, including also the case studied in~\S~\ref{sec:VALI}. We observe that the normalized cross correlation metric is appropriate to indicate the level of corruption in the acquisitions, with less consistent and more blurred information gradually observed from left (lowest corruption) to right (highest corruption).

\figstaastmine
\renewcommand{\pamine}{50}
\capsepmine
\setlength{\hcolw}{0.17\textwidth}
\centerline{
\begin{overpic}[width=\hcolw,trim={0cm 3cm 0cm 3cm},clip]{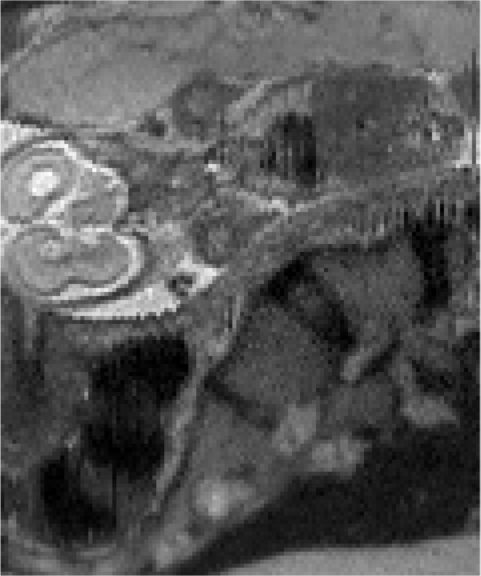}\end{overpic}
\begin{overpic}[width=\hcolw,trim={0cm 3cm 0cm 3cm},clip]{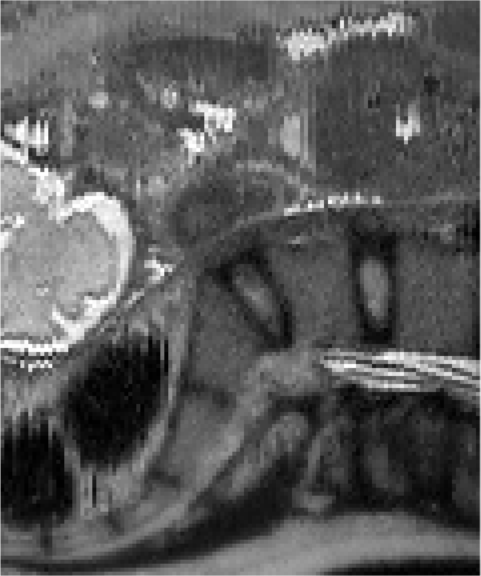}\end{overpic}
\begin{overpic}[width=\hcolw,trim={0cm 3cm 0cm 3cm},clip]{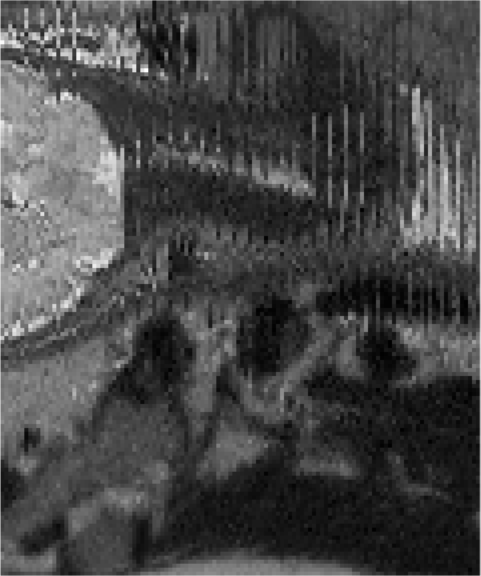}\end{overpic}
\begin{overpic}[width=\hcolw,trim={0cm 3cm 0cm 3cm},clip]{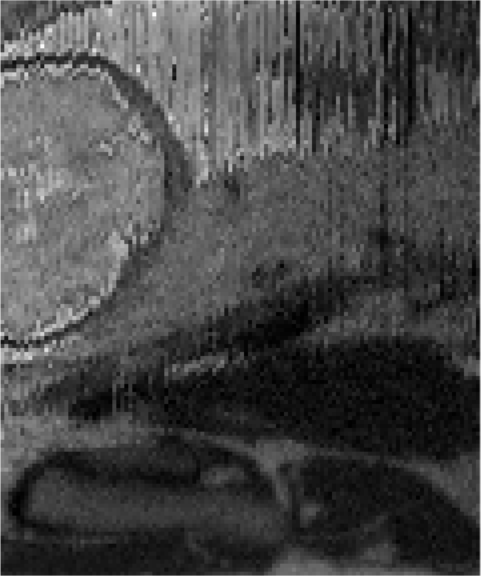}\end{overpic}
\begin{overpic}[width=\hcolw,trim={0cm 3cm 0cm 3cm},clip]{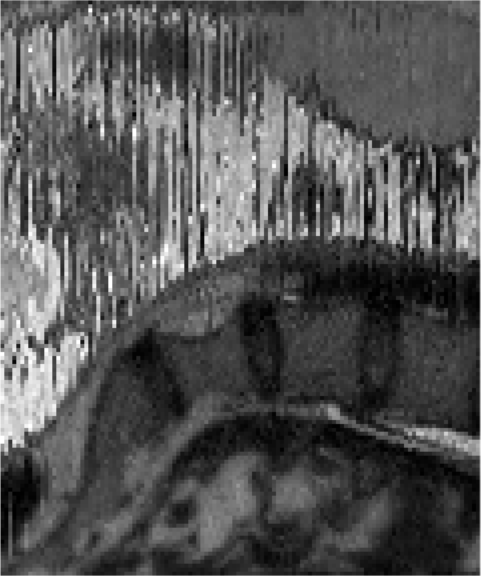}\end{overpic}\\}
\vspace{1mm}
\centerline{
\begin{overpic}[width=\hcolw,trim={0cm 3cm 0cm 3cm},clip]{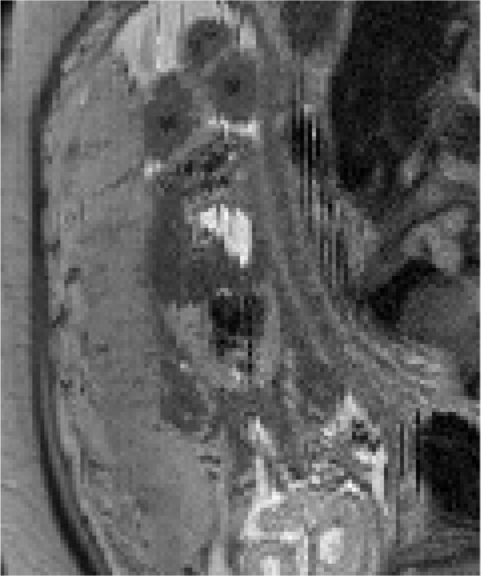}\end{overpic}
\begin{overpic}[width=\hcolw,trim={0cm 3cm 0cm 3cm},clip]{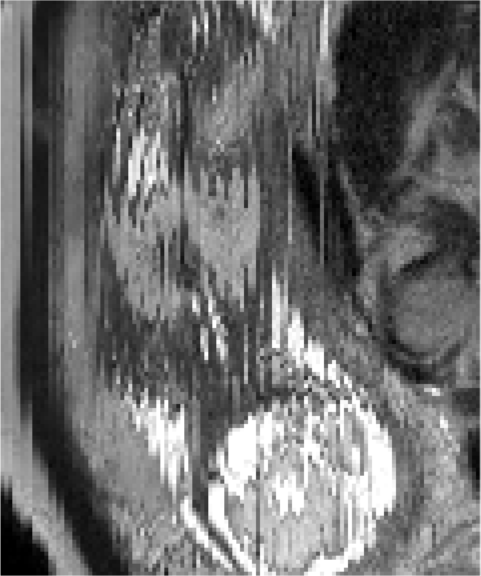}\end{overpic}
\begin{overpic}[width=\hcolw,trim={0cm 3cm 0cm 3cm},clip]{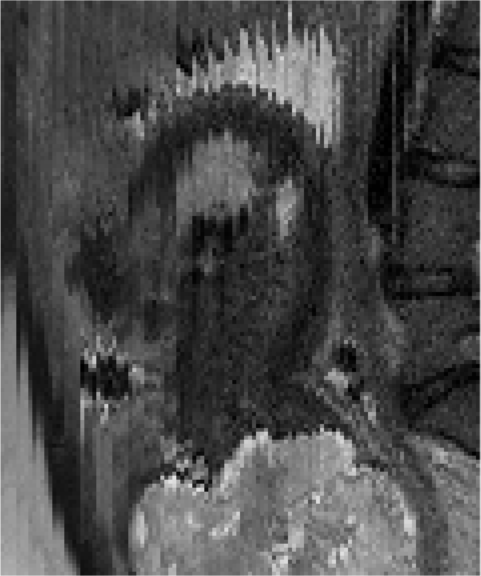}\end{overpic}
\begin{overpic}[width=\hcolw,trim={0cm 3cm 0cm 3cm},clip]{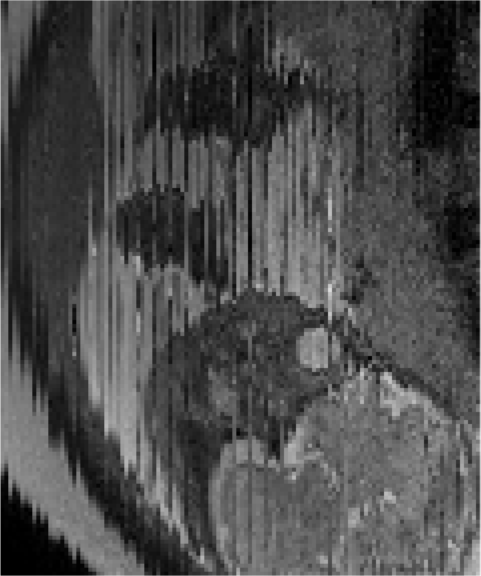}\end{overpic}
\begin{overpic}[width=\hcolw,trim={0cm 3cm 0cm 3cm},clip]{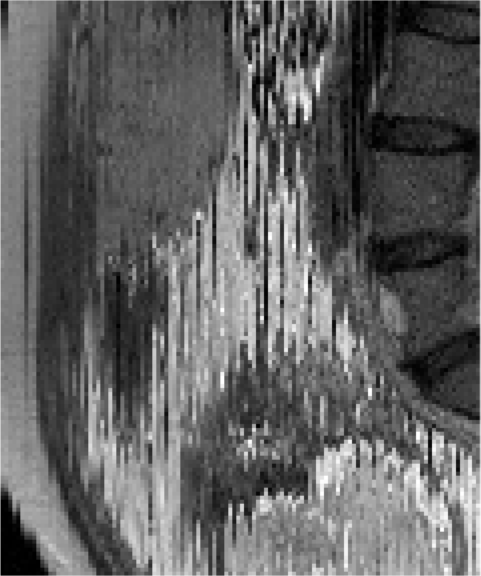}\end{overpic}\\}
\vspace{1mm}
\centerline{
\begin{overpic}[width=\hcolw,trim={0cm 3cm 0cm 3cm},clip]{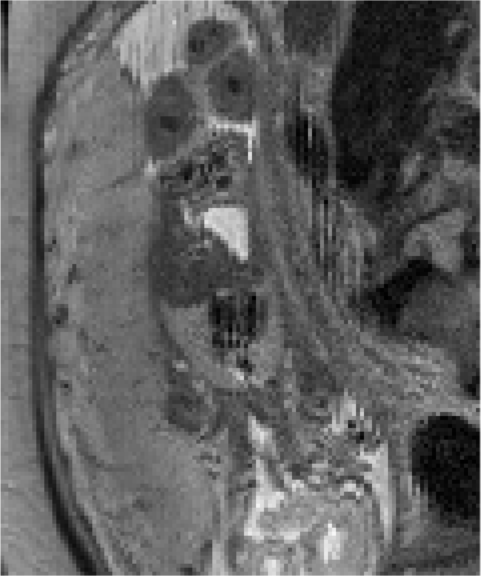}\end{overpic}
\begin{overpic}[width=\hcolw,trim={0cm 3cm 0cm 3cm},clip]{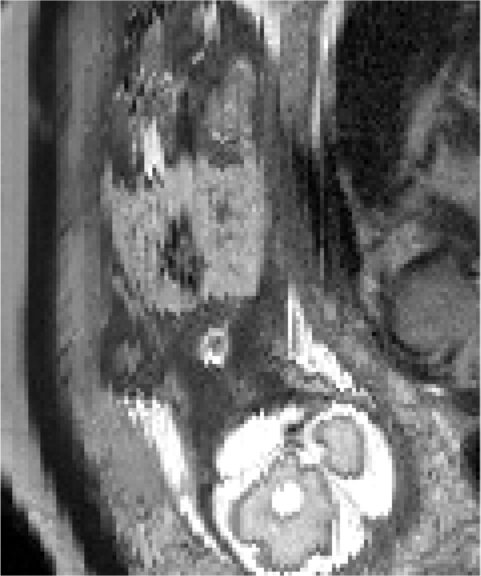}\end{overpic}
\begin{overpic}[width=\hcolw,trim={0cm 3cm 0cm 3cm},clip]{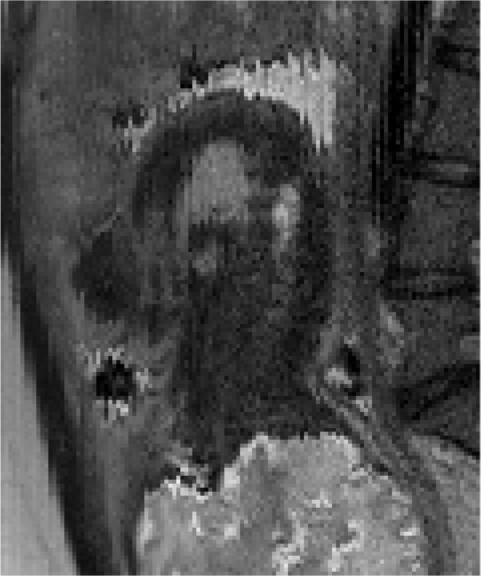}\end{overpic}
\begin{overpic}[width=\hcolw,trim={0cm 3cm 0cm 3cm},clip]{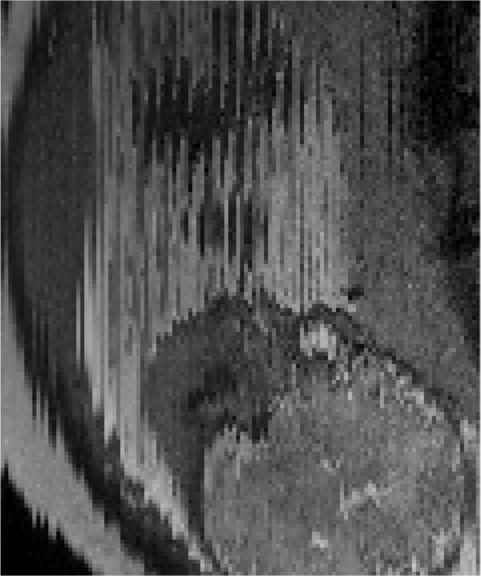}\end{overpic}
\begin{overpic}[width=\hcolw,trim={0cm 3cm 0cm 3cm},clip]{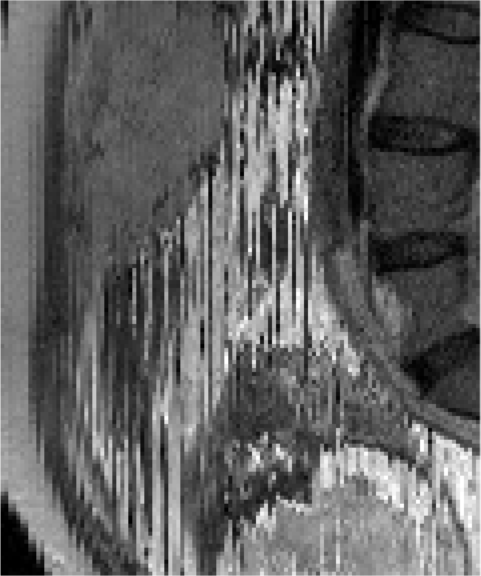}\end{overpic}\\}
\vspace{1mm}
\centerline{
\begin{overpic}[width=\hcolw,trim={0cm 3cm 0cm 3cm},clip]{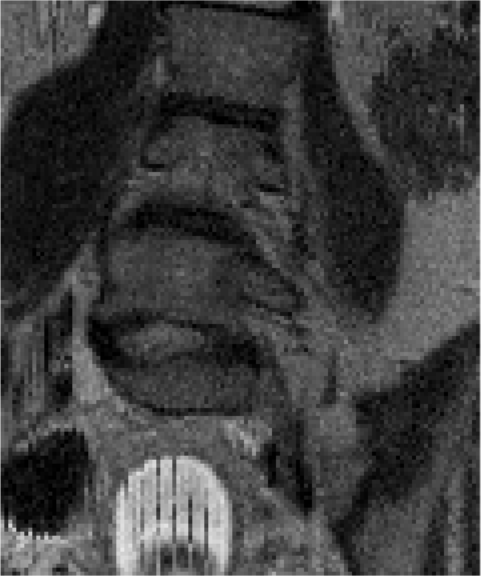}\end{overpic}
\begin{overpic}[width=\hcolw,trim={0cm 3cm 0cm 3cm},clip]{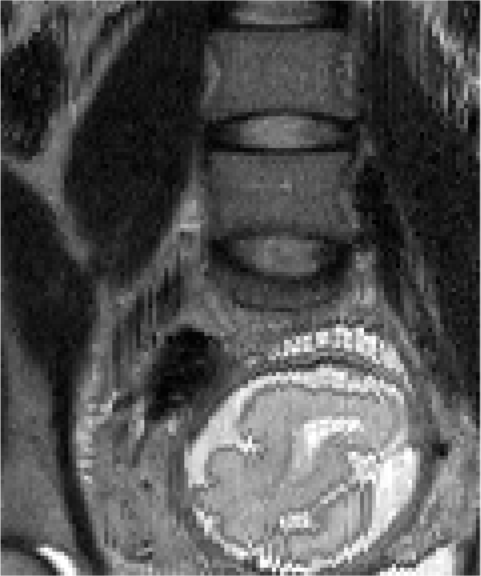}\end{overpic}
\begin{overpic}[width=\hcolw,trim={0cm 3cm 0cm 3cm},clip]{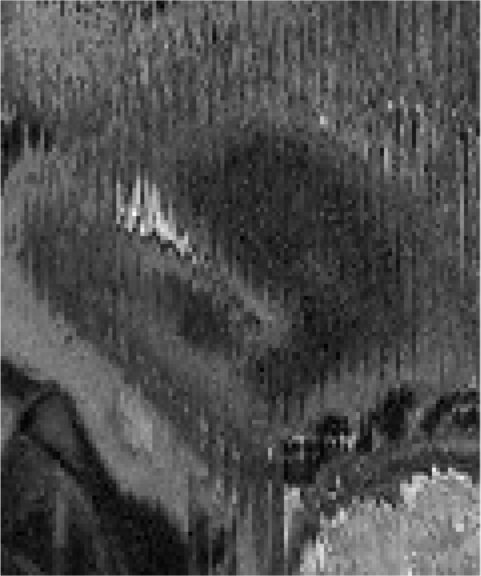}\end{overpic}
\begin{overpic}[width=\hcolw,trim={0cm 3cm 0cm 3cm},clip]{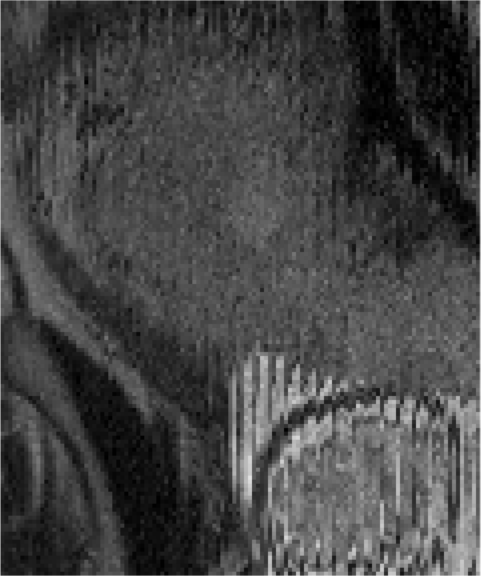}\end{overpic}
\begin{overpic}[width=\hcolw,trim={0cm 3cm 0cm 3cm},clip]{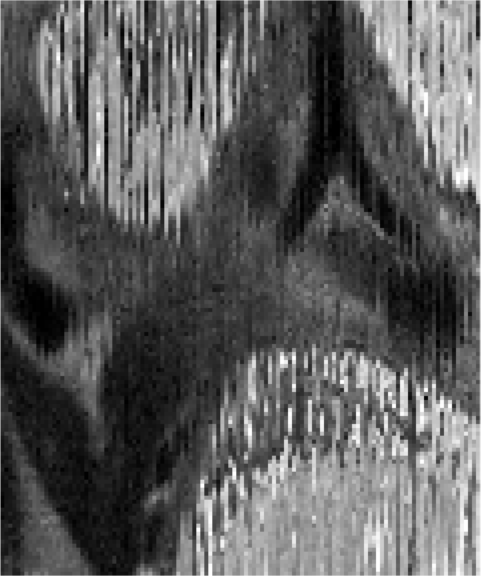}\end{overpic}\\}
\vspace{1mm}
\centerline{
\begin{overpic}[width=\hcolw,trim={0cm 3cm 0cm 3cm},clip]{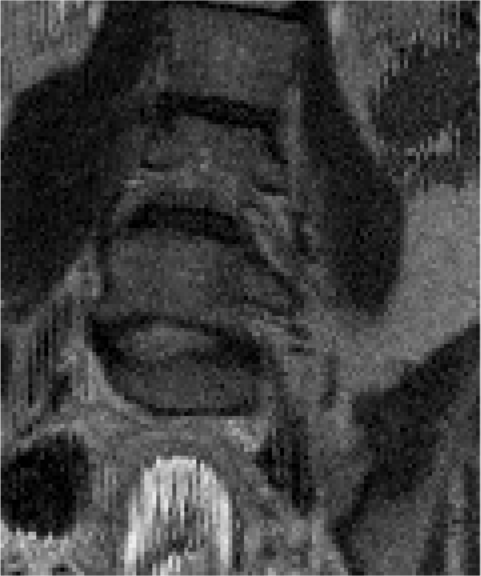}\put(\pamine,\pbmine)
{\makebox(-90,-2){\auxminebis\textbf{a)}}}\end{overpic}
\begin{overpic}[width=\hcolw,trim={0cm 3cm 0cm 3cm},clip]{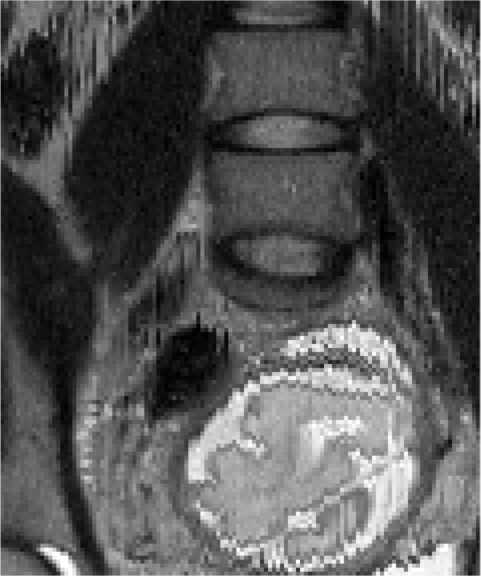}\put(\pamine,\pbmine)
{\makebox(-90,-2){\auxminebis\textbf{b)}}}\end{overpic}
\begin{overpic}[width=\hcolw,trim={0cm 3cm 0cm 3cm},clip]{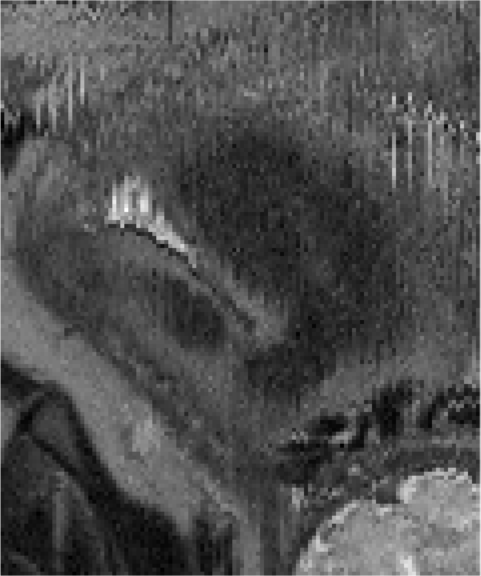}\put(\pamine,\pbmine)
{\makebox(-90,-2){\auxminebis\textbf{c)}}}\end{overpic}
\begin{overpic}[width=\hcolw,trim={0cm 3cm 0cm 3cm},clip]{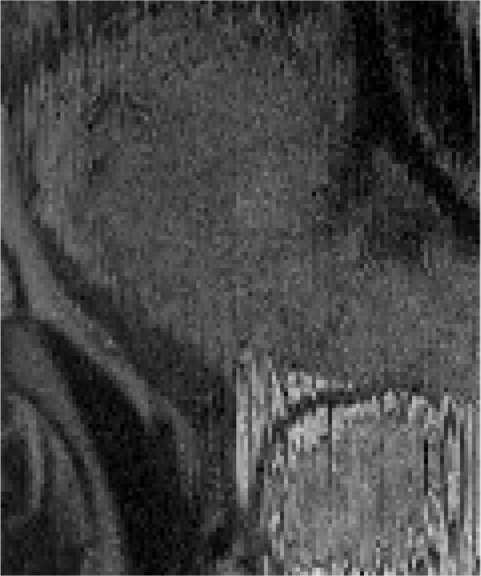}\put(\pamine,\pbmine)
{\makebox(-90,-2){\auxminebis\textbf{d)}}}\end{overpic}
\begin{overpic}[width=\hcolw,trim={0cm 3cm 0cm 3cm},clip]{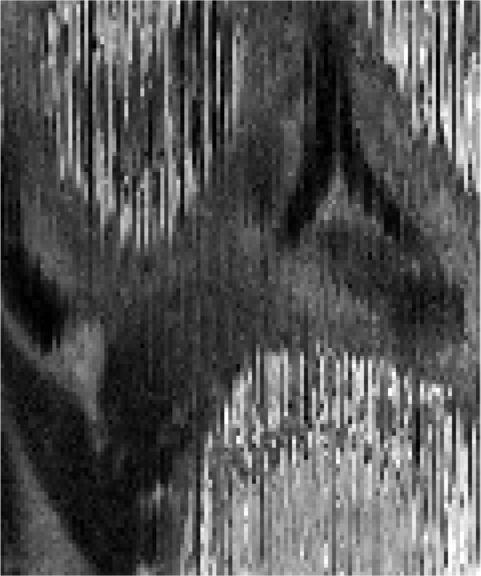}\put(\pamine,\pbmine)
{\makebox(-90,-2){\auxminebis\textbf{e)}}}\end{overpic}\\}
\vspace{1mm}
\caption{Examples of original stacks (from top to bottom respectively axial, first and second coronal repeat, and first and second sagittal repeat) in mid-planes along the readout direction for scans with \textbf{(a)} lowest, \textbf{(b)} $25$-percentile, \textbf{(c)} median, \textbf{(d)} $75$-percentile, and \textbf{(e)} highest data corruption as assessed by the normalized cross correlation between adjacent slices.}
\label{fig:FIGI}
\figendastmine

In Figs.~\ref{fig:FIGII}-\ref{fig:FIGV} we visually compare the reconstruction alternatives in~\S~\ref{sec:VALI} for the aforementioned cases, lowest (Fig.~\ref{fig:FIGII}), $25$-percentile (Fig.~\ref{fig:FIGIII}), $75$-percentile (Fig.~\ref{fig:FIGIV}), and highest degradation (Fig.~\ref{fig:FIGV}). As for the case with lowest degradation (Fig.~\ref{fig:FIGII}), we observe better resolution preservation when using the DD regularizer (Fig.~\ref{fig:FIGII}a) as compared to the handcrafted regularizer (Fig.~\ref{fig:FIGII}c), for instance when looking at the brain cortex. Blurring is also noticeable without motion compensation (Fig.~\ref{fig:FIGII}e) despite this is a case with minimum fetal motion. An example of the impact of non-suppressed artifacts in non-robust reconstructions (Fig.~\ref{fig:FIGII}d) can be observed in the area within the blue ellipse, where the bright stripe disrupting the geometry of the hand is suppressed when the robust formulation is adopted.

\ifdefined\tmiformat
	\newpage
\fi

Interpretation of results is not very different as degradation of acquired data becomes more prominent. We note that as motion grows, the regularization has a stronger impact in registration (see for instance the area enclosed in blue in Fig.~\ref{fig:FIGIV}a versus Fig.~\ref{fig:FIGIV}b). As for non-robust reconstructions, subtle localized differences can be observed at different locations in the provided snapshots, for instance in the muscle regions enclosed in blue in Fig.~\ref{fig:FIGIII}. However, these become more prominent for highly corrupted datasets, as in the mouth area in blue in Fig.~\ref{fig:FIGV}, with stronger spurious features in Fig.~\ref{fig:FIGV}d when compared to Fig.~\ref{fig:FIGV}a. Impact of motion correction also becomes gradually more important as degradation becomes more pronounced. We observe stronger contrast improvements when comparing the brain area of Fig.~\ref{fig:FIGIII}a versus that of Fig.~\ref{fig:FIGIII}e than those observed in Fig.~\ref{fig:FIGII}. As we move towards most degraded cases, more and more structures are resolved by motion correction. For instance, within the green ellipse in Fig.~\ref{fig:FIGV}, we observe noticeably improved delineation of the organs in the thoracic cavity, although many fine detailed structures are missing in this maximally degraded scan.

\figstaastmine
\renewcommand{\pamine}{50}
\ifdefined\tmiformat
	\setlength{\hcolw}{0.19\textwidth}
\else
	\setlength{\hcolw}{0.17\textwidth}
\fi	
\centerline{
\begin{overpic}[width=\hcolw]{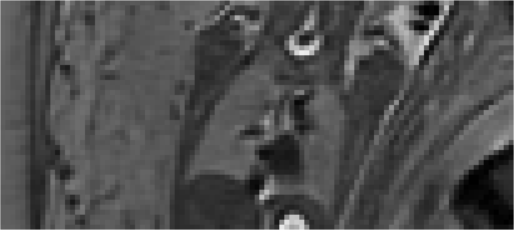}\end{overpic}
\begin{overpic}[width=\hcolw]{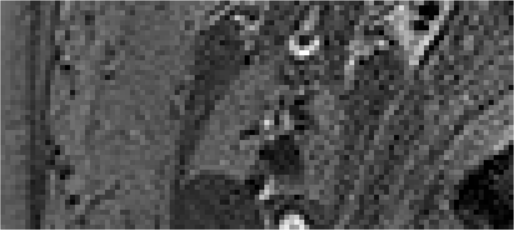}\end{overpic}
\begin{overpic}[width=\hcolw]{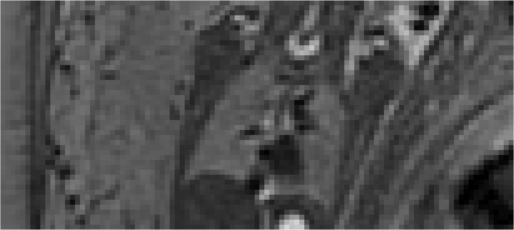}\end{overpic}
\begin{overpic}[width=\hcolw]{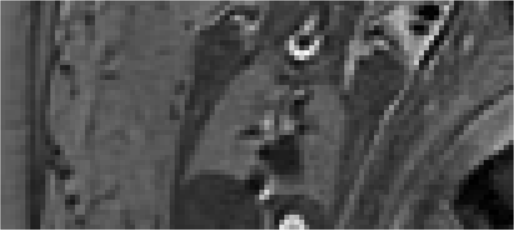}\end{overpic}
\begin{overpic}[width=\hcolw]{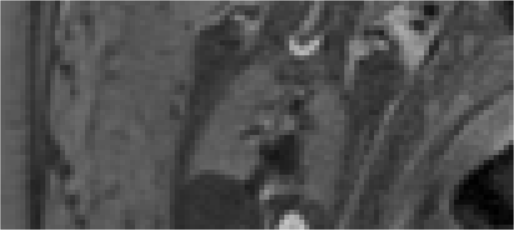}\end{overpic}\\}
\vspace{1mm}
\centerline{
\begin{overpic}[width=\hcolw]{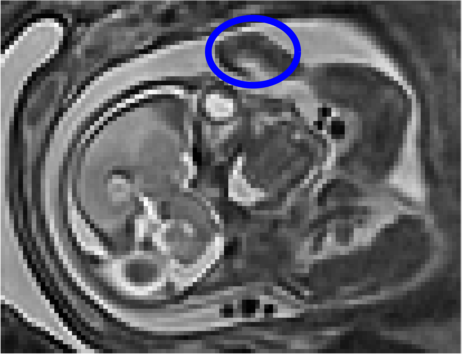}\end{overpic}
\begin{overpic}[width=\hcolw]{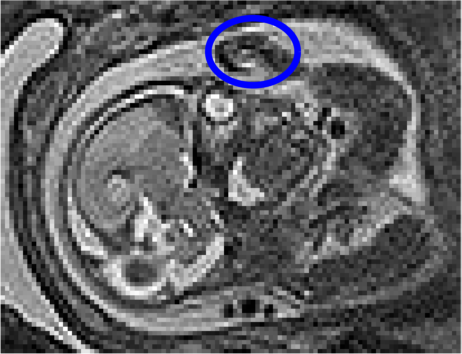}\end{overpic}
\begin{overpic}[width=\hcolw]{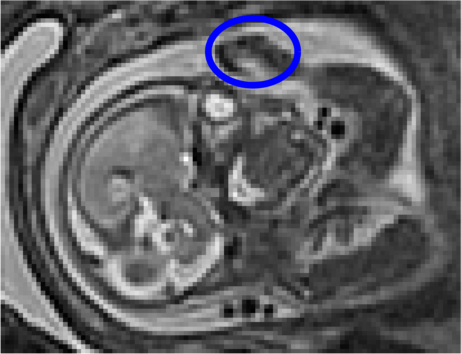}\end{overpic}
\begin{overpic}[width=\hcolw]{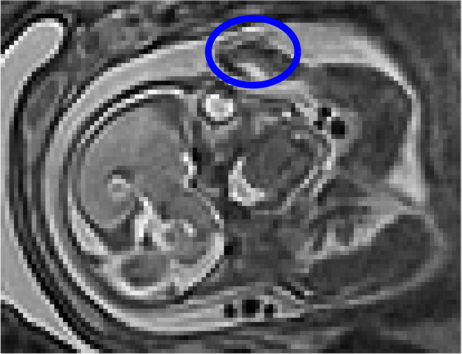}\end{overpic}
\begin{overpic}[width=\hcolw]{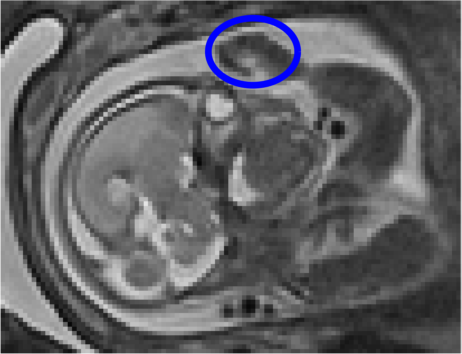}\end{overpic}\\}
\vspace{1mm}
\centerline{
\begin{overpic}[width=\hcolw]{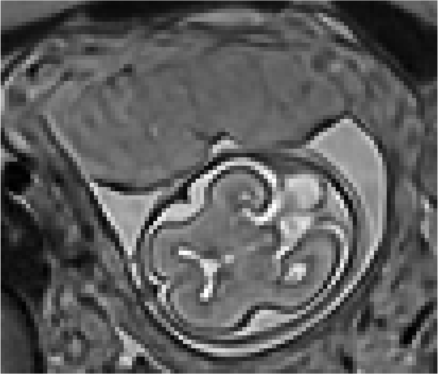}\put(\pamine,\pbmine)
{\makebox(-90,-2){\auxminebis\textbf{a)}}}\end{overpic}
\begin{overpic}[width=\hcolw]{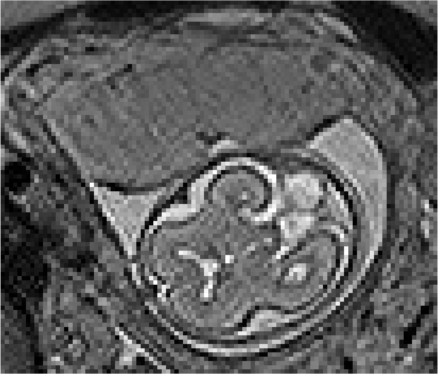}\put(\pamine,\pbmine)
{\makebox(-90,-2){\auxminebis\textbf{b)}}}\end{overpic}
\begin{overpic}[width=\hcolw]{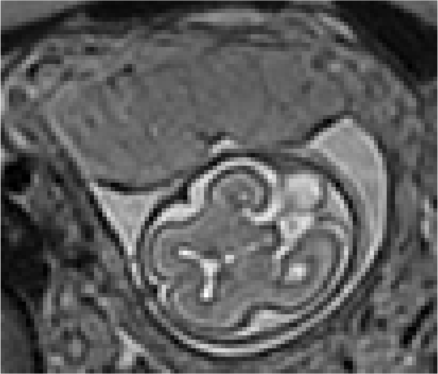}\put(\pamine,\pbmine)
{\makebox(-90,-2){\auxminebis\textbf{c)}}}\end{overpic}
\begin{overpic}[width=\hcolw]{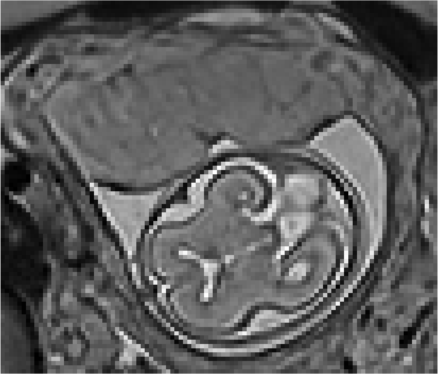}\put(\pamine,\pbmine)
{\makebox(-90,-2){\auxminebis\textbf{d)}}}\end{overpic}
\begin{overpic}[width=\hcolw]{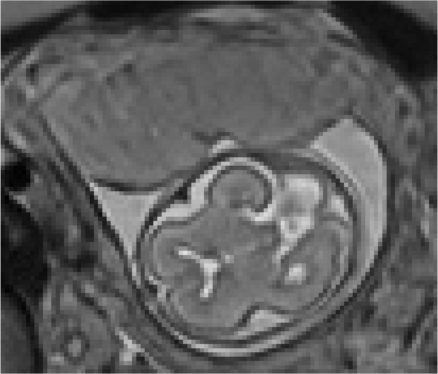}\put(\pamine,\pbmine)
{\makebox(-90,-2){\auxminebis\textbf{e)}}}\end{overpic}\\
}
\capsepmine
\caption{Comparison of different reconstruction alternatives in the case with lowest degradation. From top to bottom, coronal, sagittal and axial planes in the mother's geometry. Reconstructions \textbf{(a)} based on the full model ($\mathbf{x}$); \textbf{(b)} without regularization ($\mathbf{x}_{\lambda=0}$); \textbf{(c)} with handcrafted regularization ($\mathbf{x}_{\mathcal{H}}$); \textbf{(d)} without the robust formulation ($\mathbf{x}_{\tau=0}$); \textbf{(e)} without motion correction ($\mathbf{x}_{\sigma_{\text{I}}\to\infty}$). The blue ellipse highlights local artifacts, noise or blurring in the hand when taking off any of the main components of our formulation.}
\label{fig:FIGII}
\figendastmine

\figstaastmine
\renewcommand{\pamine}{50}
\ifdefined\tmiformat
	\setlength{\hcolw}{0.19\textwidth}
\else
	\setlength{\hcolw}{0.17\textwidth}
\fi	
\centerline{
\begin{overpic}[width=\hcolw]{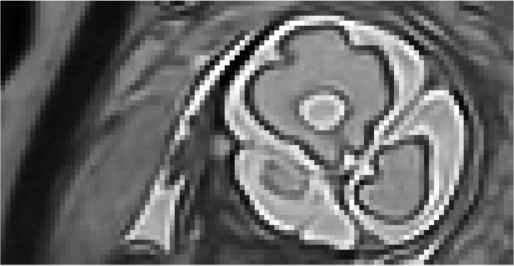}\end{overpic}
\begin{overpic}[width=\hcolw]{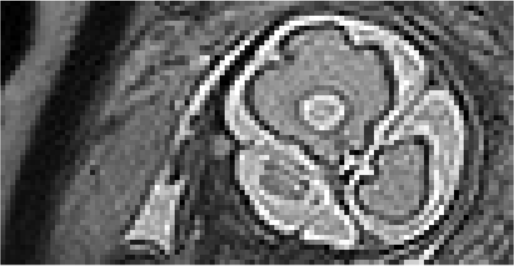}\end{overpic}
\begin{overpic}[width=\hcolw]{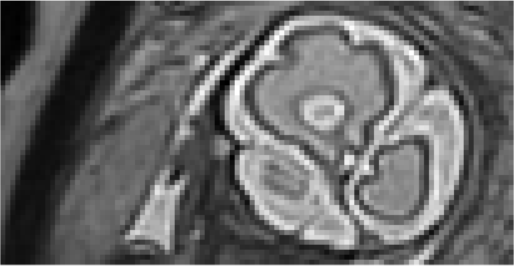}\end{overpic}
\begin{overpic}[width=\hcolw]{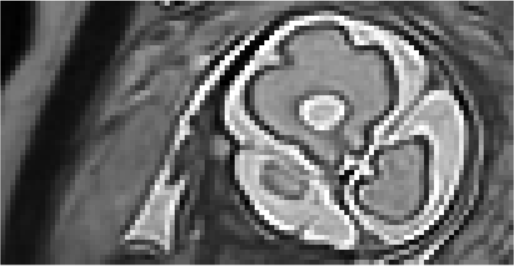}\end{overpic}
\begin{overpic}[width=\hcolw]{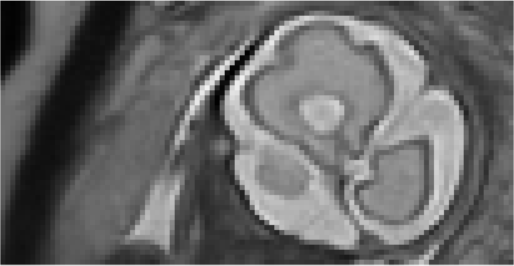}\end{overpic}\\}
\vspace{1mm}
\centerline{
\begin{overpic}[width=\hcolw]{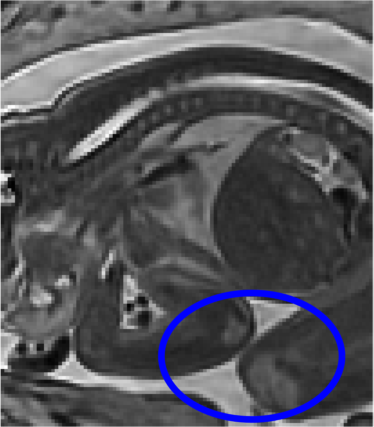}\end{overpic}
\begin{overpic}[width=\hcolw]{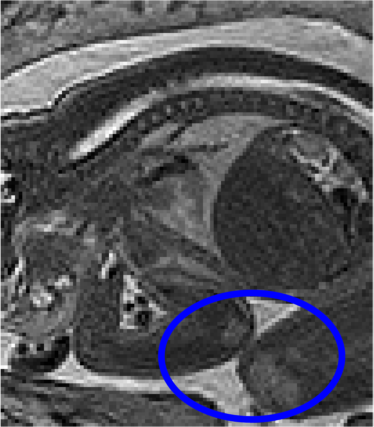}\end{overpic}
\begin{overpic}[width=\hcolw]{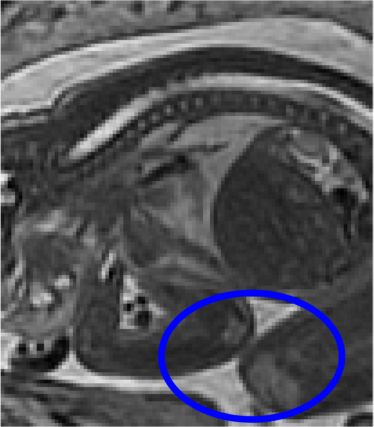}\end{overpic}
\begin{overpic}[width=\hcolw]{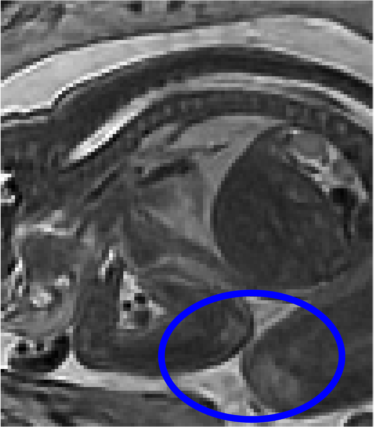}\end{overpic}
\begin{overpic}[width=\hcolw]{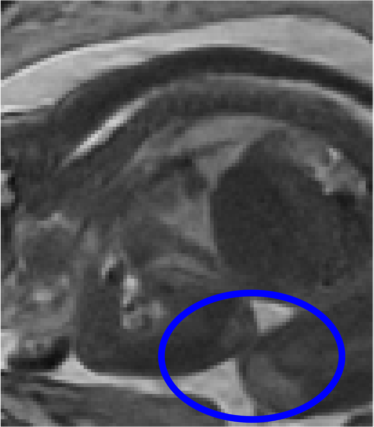}\end{overpic}\\}
\vspace{1mm}
\centerline{
\begin{overpic}[width=\hcolw]{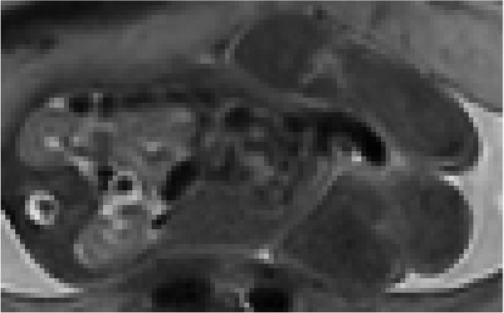}\put(\pamine,\pbmine)
{\makebox(-90,-2){\auxminebis\textbf{a)}}}\end{overpic}
\begin{overpic}[width=\hcolw]{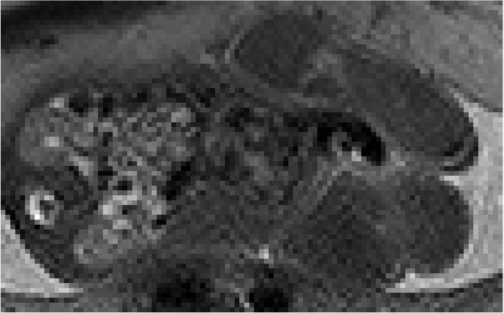}\put(\pamine,\pbmine)
{\makebox(-90,-2){\auxminebis\textbf{b)}}}\end{overpic}
\begin{overpic}[width=\hcolw]{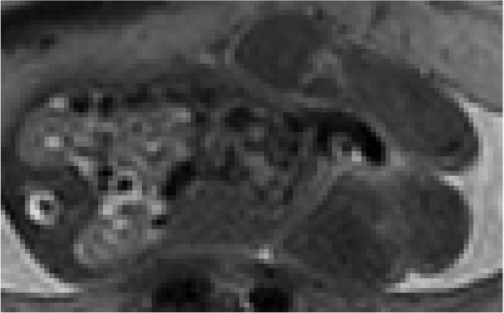}\put(\pamine,\pbmine)
{\makebox(-90,-2){\auxminebis\textbf{c)}}}\end{overpic}
\begin{overpic}[width=\hcolw]{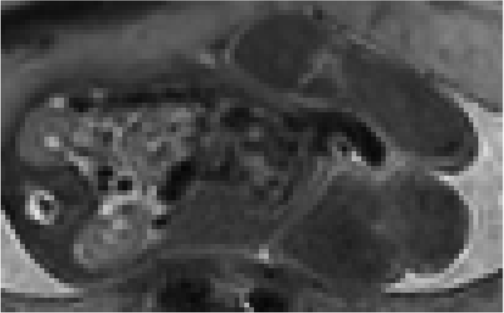}\put(\pamine,\pbmine)
{\makebox(-90,-2){\auxminebis\textbf{d)}}}\end{overpic}
\begin{overpic}[width=\hcolw]{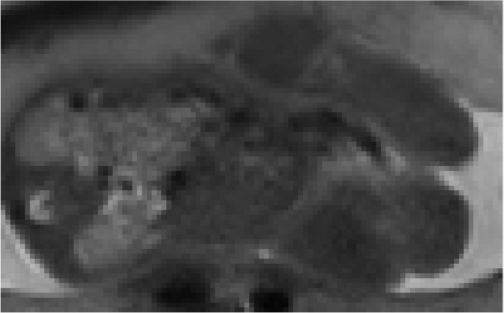}\put(\pamine,\pbmine)
{\makebox(-90,-2){\auxminebis\textbf{e)}}}\end{overpic}\\
}
\capsepmine
\caption{Comparison of different reconstruction alternatives in the case with $25$-percentile degradation. From top to bottom, coronal, sagittal and axial planes in the mother's geometry. Reconstructions \textbf{(a)} based on the full model ($\mathbf{x}$); \textbf{(b)} without regularization ($\mathbf{x}_{\lambda=0}$); \textbf{(c)} with handcrafted regularization ($\mathbf{x}_{\mathcal{H}}$); \textbf{(d)} without the robust formulation ($\mathbf{x}_{\tau=0}$); \textbf{(e)} without motion correction ($\mathbf{x}_{\sigma_{\text{I}}\to\infty}$). The blue ellipse highlights most consistent information within the muscular regions when adopting the full formulation.}
\label{fig:FIGIII}
\figendastmine

\figstaastmine
\renewcommand{\pamine}{50}
\ifdefined\tmiformat
	\setlength{\hcolw}{0.19\textwidth}
\else
	\setlength{\hcolw}{0.17\textwidth}
\fi	
\centerline{
\begin{overpic}[width=\hcolw]{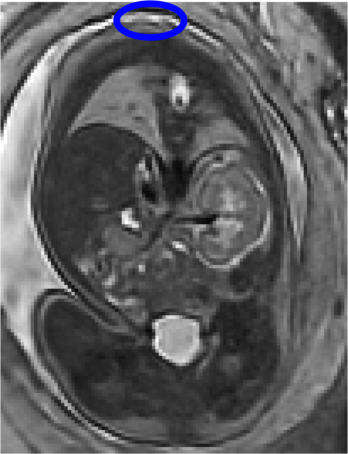}\end{overpic}
\begin{overpic}[width=\hcolw]{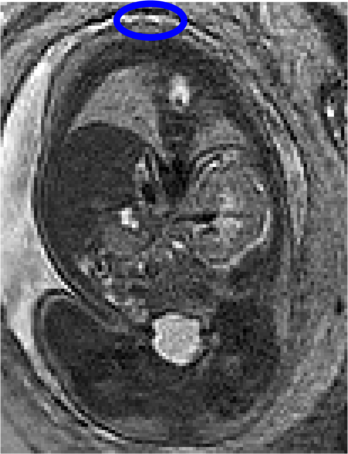}\end{overpic}
\begin{overpic}[width=\hcolw]{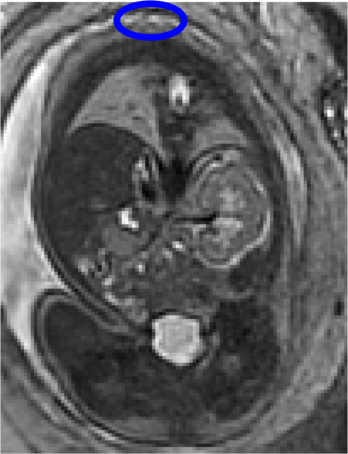}\end{overpic}
\begin{overpic}[width=\hcolw]{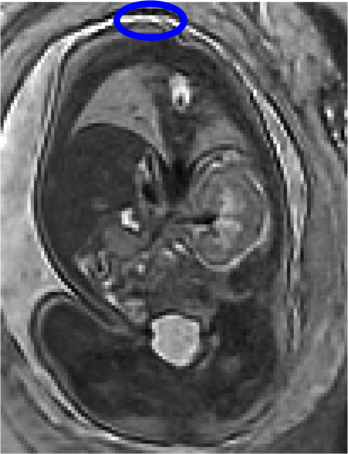}\end{overpic}
\begin{overpic}[width=\hcolw]{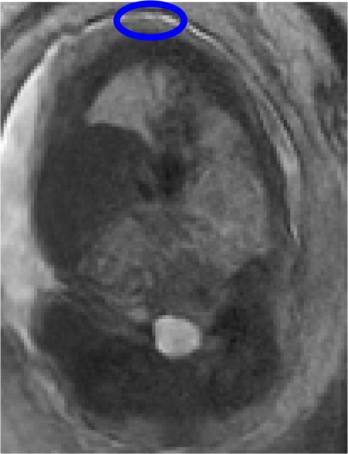}\end{overpic}\\}
\vspace{1mm}
\centerline{
\begin{overpic}[width=\hcolw]{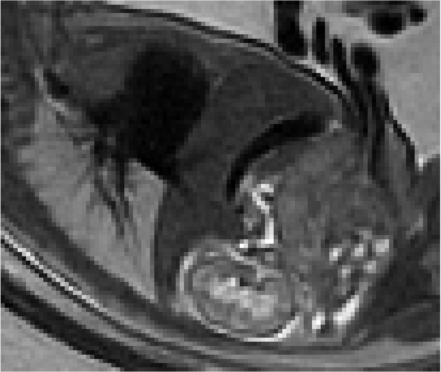}\end{overpic}
\begin{overpic}[width=\hcolw]{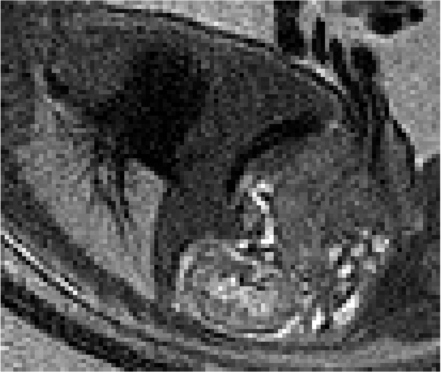}\end{overpic}
\begin{overpic}[width=\hcolw]{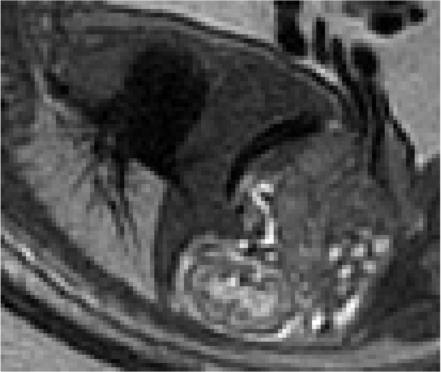}\end{overpic}
\begin{overpic}[width=\hcolw]{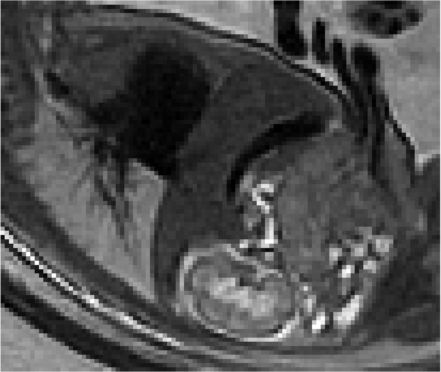}\end{overpic}
\begin{overpic}[width=\hcolw]{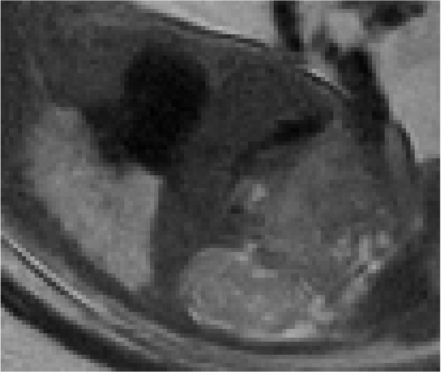}\end{overpic}\\}
\vspace{1mm}
\centerline{
\begin{overpic}[width=\hcolw]{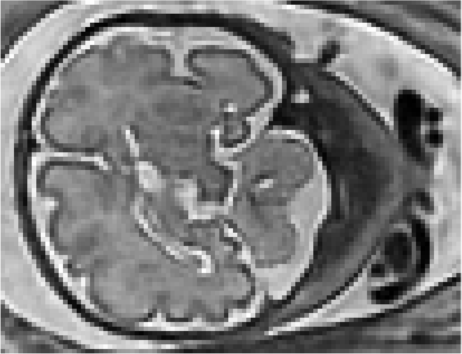}\put(\pamine,\pbmine)
{\makebox(-90,-2){\auxminebis\textbf{a)}}}\end{overpic}
\begin{overpic}[width=\hcolw]{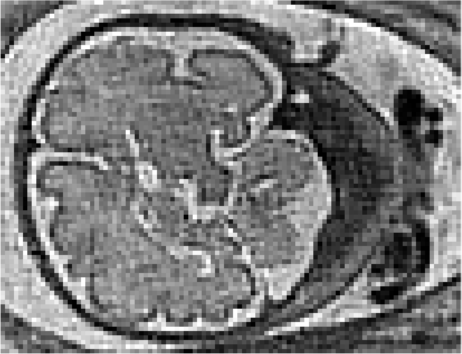}\put(\pamine,\pbmine)
{\makebox(-90,-2){\auxminebis\textbf{b)}}}\end{overpic}
\begin{overpic}[width=\hcolw]{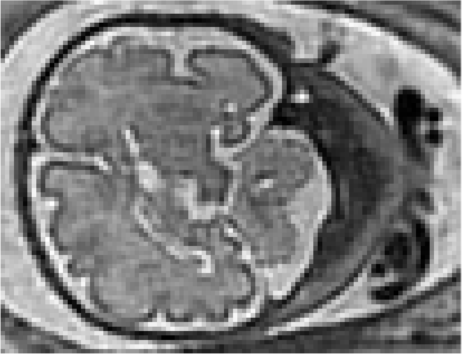}\put(\pamine,\pbmine)
{\makebox(-90,-2){\auxminebis\textbf{c)}}}\end{overpic}
\begin{overpic}[width=\hcolw]{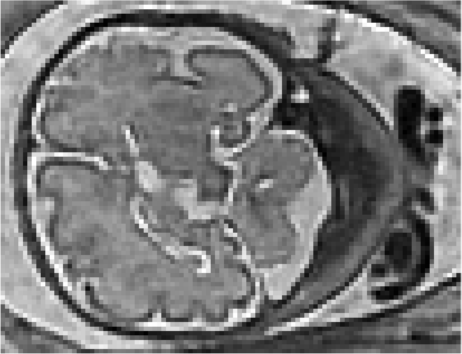}\put(\pamine,\pbmine)
{\makebox(-90,-2){\auxminebis\textbf{d)}}}\end{overpic}
\begin{overpic}[width=\hcolw]{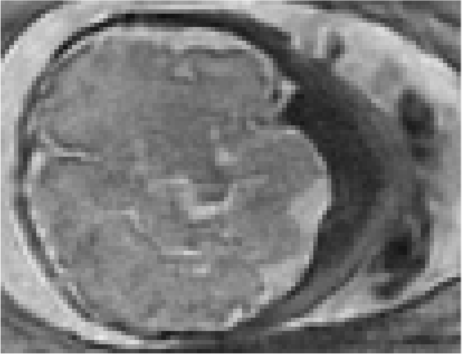}\put(\pamine,\pbmine)
{\makebox(-90,-2){\auxminebis\textbf{e)}}}\end{overpic}\\
}
\capsepmine
\caption{Comparison of different reconstruction alternatives in the case with $75$-percentile degradation. From top to bottom, coronal, sagittal and axial planes in the mother's geometry. Reconstructions \textbf{(a)} based on the full model ($\mathbf{x}$); \textbf{(b)} without regularization ($\mathbf{x}_{\lambda=0}$); \textbf{(c)} with handcrafted regularization ($\mathbf{x}_{\mathcal{H}}$); \textbf{(d)} without the robust formulation ($\mathbf{x}_{\tau=0}$); \textbf{(e)} without motion correction ($\mathbf{x}_{\sigma_{\text{I}}\to\infty}$). The blue ellipse highlights potential slice matching improvements when using the DD instead of the handcrafted regularizer.}
\label{fig:FIGIV}
\figendastmine

\figstaastmine
\renewcommand{\pamine}{50}
\ifdefined\tmiformat
	\setlength{\hcolw}{0.19\textwidth}
\else
	\setlength{\hcolw}{0.17\textwidth}
\fi	
\centerline{
\begin{overpic}[width=\hcolw]{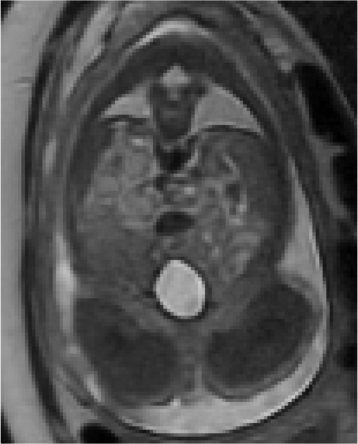}\end{overpic}
\begin{overpic}[width=\hcolw]{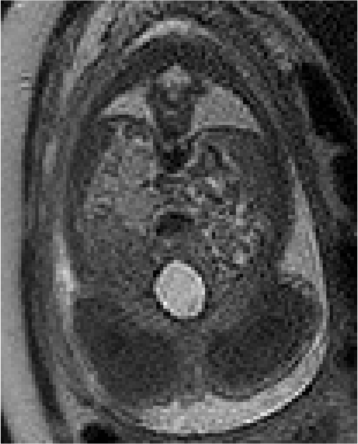}\end{overpic}
\begin{overpic}[width=\hcolw]{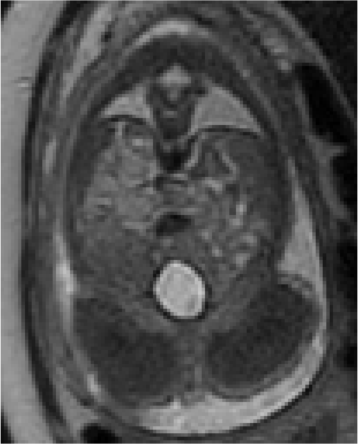}\end{overpic}
\begin{overpic}[width=\hcolw]{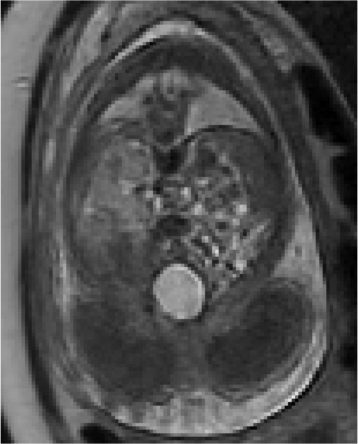}\end{overpic}
\begin{overpic}[width=\hcolw]{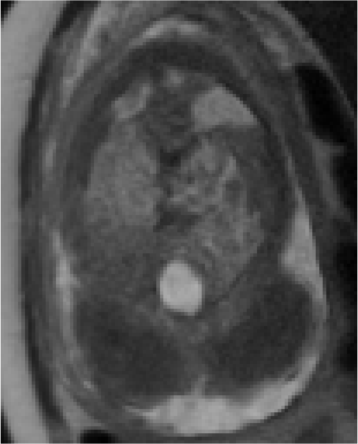}\end{overpic}\\}
\vspace{1mm}
\centerline{
\begin{overpic}[width=\hcolw]{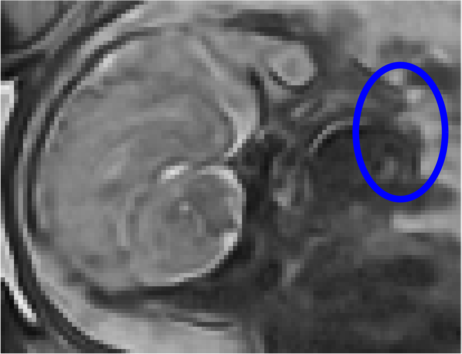}\end{overpic}
\begin{overpic}[width=\hcolw]{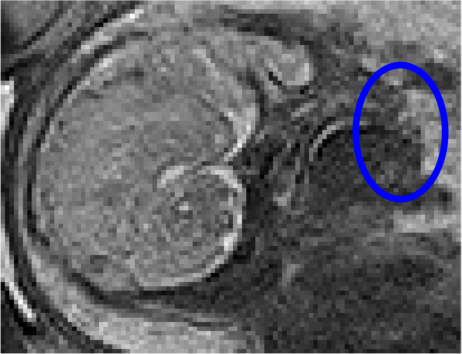}\end{overpic}
\begin{overpic}[width=\hcolw]{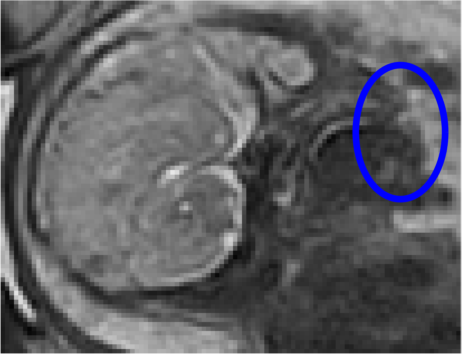}\end{overpic}
\begin{overpic}[width=\hcolw]{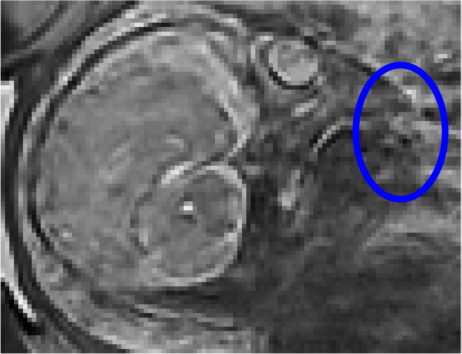}\end{overpic}
\begin{overpic}[width=\hcolw]{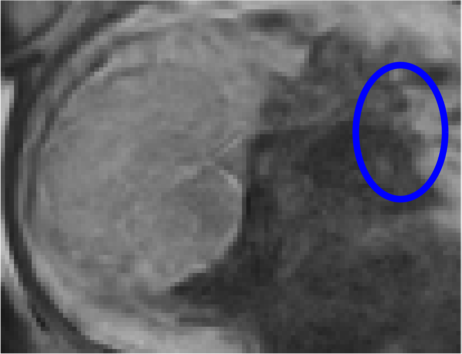}\end{overpic}\\}
\vspace{1mm}
\centerline{
\begin{overpic}[width=\hcolw]{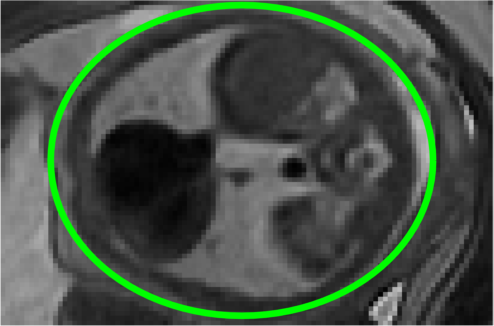}\put(\pamine,\pbmine)
{\makebox(-90,-2){\auxminebis\textbf{a)}}}\end{overpic}
\begin{overpic}[width=\hcolw]{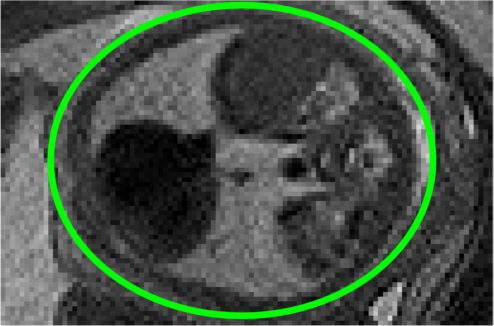}\put(\pamine,\pbmine)
{\makebox(-90,-2){\auxminebis\textbf{b)}}}\end{overpic}
\begin{overpic}[width=\hcolw]{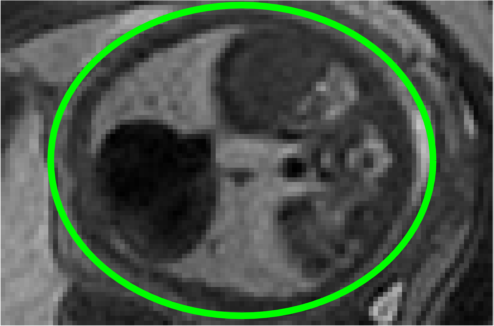}\put(\pamine,\pbmine)
{\makebox(-90,-2){\auxminebis\textbf{c)}}}\end{overpic}
\begin{overpic}[width=\hcolw]{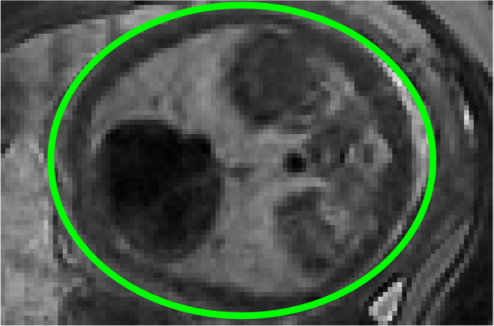}\put(\pamine,\pbmine)
{\makebox(-90,-2){\auxminebis\textbf{d)}}}\end{overpic}
\begin{overpic}[width=\hcolw]{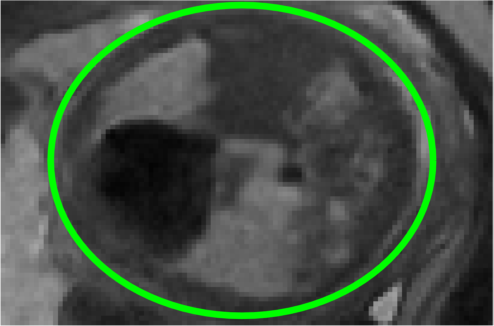}\put(\pamine,\pbmine)
{\makebox(-90,-2){\auxminebis\textbf{e)}}}\end{overpic}\\
}
\capsepmine
\caption{Comparison of different reconstruction alternatives in the case with highest degradation. From top to bottom, coronal, sagittal and axial planes in the mother's geometry. Reconstructions \textbf{(a)} based on the full model ($\mathbf{x}$); \textbf{(b)} without regularization ($\mathbf{x}_{\lambda=0}$); \textbf{(c)} with handcrafted regularization ($\mathbf{x}_{\mathcal{H}}$); \textbf{(d)} without the robust formulation ($\mathbf{x}_{\tau=0}$); \textbf{(e)} without motion correction ($\mathbf{x}_{\sigma_{\text{I}}\to\infty}$). The blue ellipse highlights smallest level of artifacts in the mouth when using the full formulation. The green ellipse highlights the ability to resolve the thoracic cavity for maximally degraded input data.}
\label{fig:FIGV}
\figendastmine

In Fig.~\ref{fig:FIGVI}, we show PSD curves analogous to those of Fig.~\ref{fig:FIG05}a of the main manuscript for the selected cases with different levels of acquired data corruption. Results confirm interpretation given in~\S~\ref{sec:VALI}. We also observe that PSD values are generally lower in reconstructions of most degraded datasets. Additionally, the non-robust and full model reconstruction PSDs differences increase as degradation becomes bigger, from Fig.~\ref{fig:FIGVI}a, passing through Fig.~\ref{fig:FIG05}a, to Fig.~\ref{fig:FIGVI}d. This behaviour is consistent with suppression of increased levels of artifacted structures by the robust formulation.

\figstaastmine
\renewcommand{\pamine}{50}
\vspace{2mm}
\centerline{
\begin{overpic}[width=0.49\textwidth]{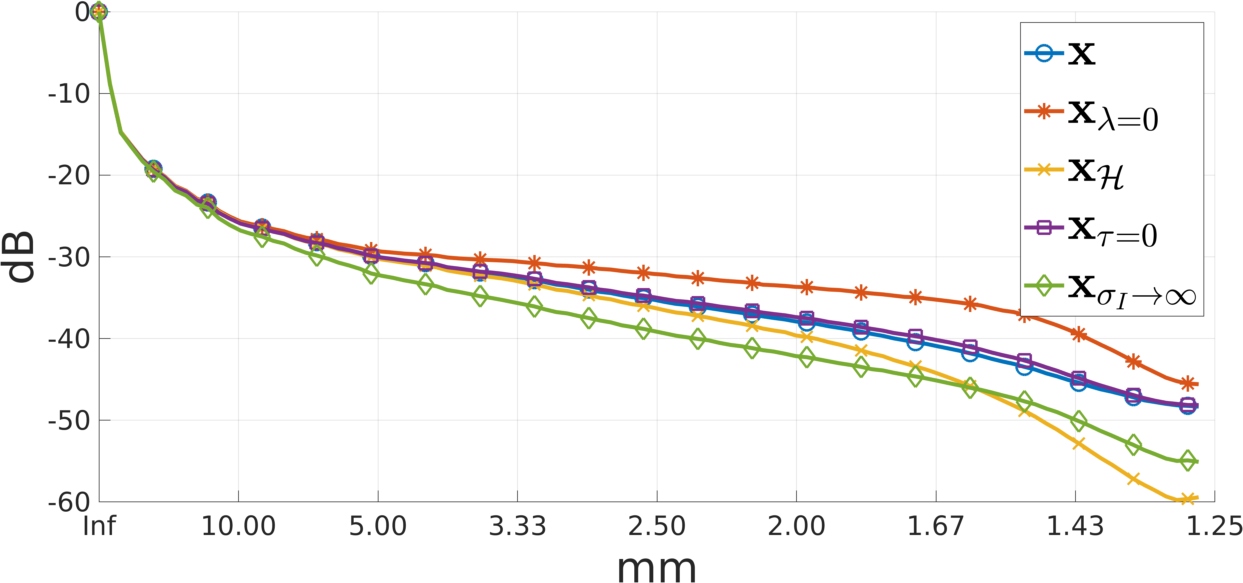}\put(\pamine,\pbmine)
{\makebox(-75,8){\auxminebis\textbf{a)}}}\end{overpic}
\begin{overpic}[width=0.49\textwidth]{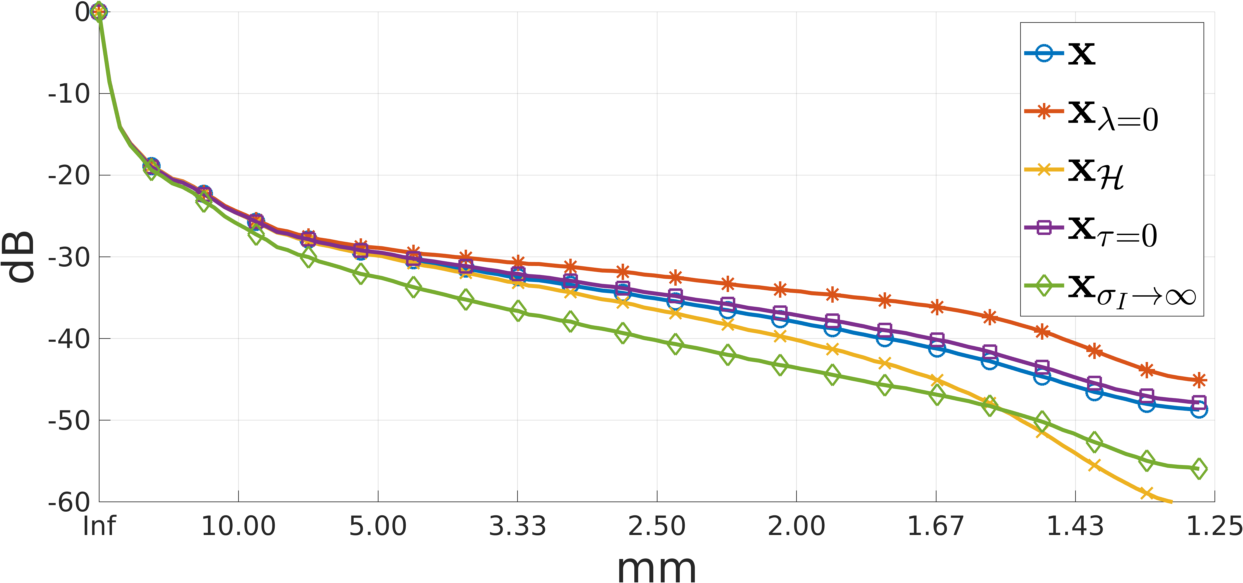}\put(\pamine,\pbmine)
{\makebox(-75,8){\auxminebis\textbf{b)}}}\end{overpic}}
\vspace{2mm}
\centerline{
\begin{overpic}[width=0.49\textwidth]{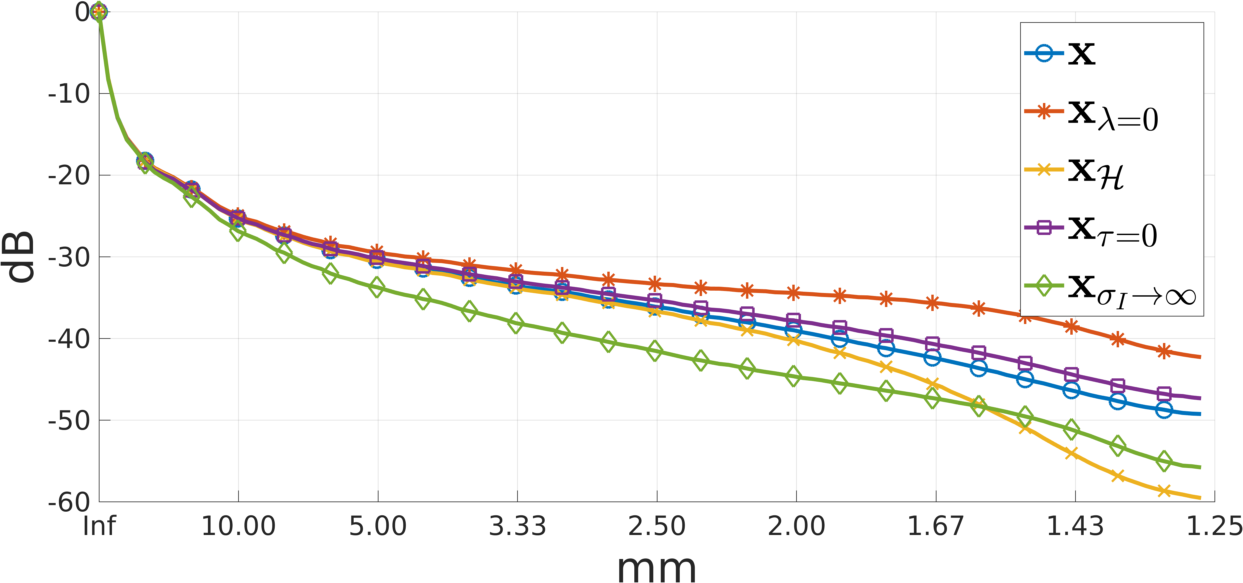}\put(\pamine,\pbmine)
{\makebox(-75,8){\auxminebis\textbf{c)}}}\end{overpic}
\begin{overpic}[width=0.49\textwidth]{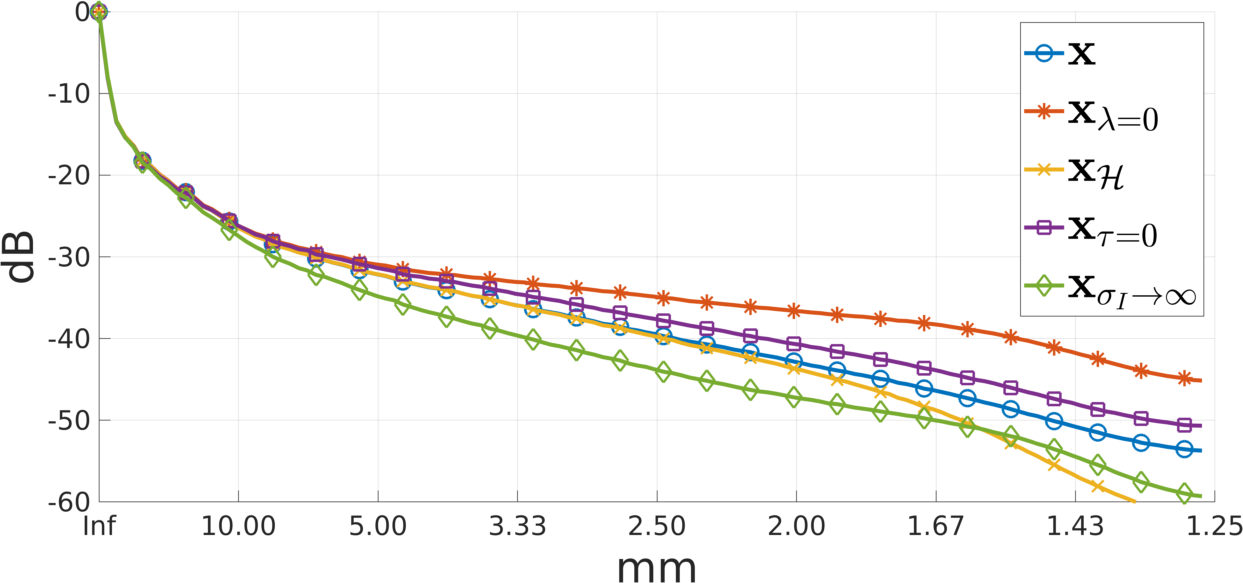}\put(\pamine,\pbmine)
{\makebox(-75,8){\auxminebis\textbf{d)}}}\end{overpic}}
\capsepmine
\caption{PSD comparisons. Studied reconstruction alternatives in the cases with \textbf{a)} lowest, \textbf{b)} $25$-percentile, \textbf{c)} $75$-percentile, and \textbf{d)} highest degradation (uterus ROI).}
\label{fig:FIGVI}
\figendastmine

\end{document}